\documentclass{aa}  

\usepackage[colorlinks]{hyperref}

\usepackage{booktabs}
\usepackage{tabularx} 
\usepackage{scrextend}
\let\orgautoref\autoref
\renewcommand{\autoref}
        {\def\equationautorefname{Eq.}%
         \def\figureautorefname{Fig.}%
         \def\sectionautorefname{Sect.}%
         \def\subsectionautorefname{Sect.}%
         \def\subsubsectionautorefname{Sect.}%
         \orgautoref}

\usepackage{xcolor}
\definecolor{dark-red}{rgb}{0.9,0.0,0.0}
\definecolor{dark-blue}{rgb}{0.15,0.15,0.9}
\definecolor{dark-green}{rgb}{0.15,0.8,0.15}
\definecolor{medium-blue}{rgb}{0,0,0.9}
\hypersetup{linkcolor={dark-blue},citecolor={dark-blue}, urlcolor={medium-blue}
}

\newcommand{\topline}{
    \hline\hline
    \noalign{\smallskip}
}
\newcommand{\midline}{
      \noalign{\smallskip}
      \hline
      \noalign{\smallskip}
}
\newcommand{\bottomline}{
    \noalign{\smallskip}
    \hline
}

\makeatletter
\renewcommand*\aa@pageof{, page \thepage{} of \pageref*{LastPage}} 
\makeatother

\usepackage{changepage} 
\usepackage{adjustbox}
\usepackage{rotating}

\usepackage{graphicx}
\usepackage{txfonts}
\usepackage{soul}

\begin{document} 

\title{A public HARPS radial velocity database corrected for systematic errors\thanks{
Based on observations collected at the European Organization for Astronomical Research in the Southern Hemisphere under ESO programmes (see {\it acknowledgements} for a full list of used programmes).}}
 

\author{
    Trifon Trifonov\inst{1,2}
    \and Lev Tal-Or\inst{3,4}
    \and Mathias Zechmeister\inst{5} 
    \and Adrian Kaminski\inst{6}
    \and Shay Zucker\inst{4} 
    \and Tsevi Mazeh\inst{7} 
}

\institute{
    Max-Planck-Institut f\"ur Astronomie, K\"onigstuhl 17, D-69117 Heidelberg, Germany\\
    \email{trifonov@mpia.de}
    \and Department
 of Astronomy, Sofia University "St Kliment Ohridski", 5 James Bourchier Blvd, BG-1164 Sofia, Bulgaria
    \and Department of Physics, Ariel University, Ariel 40700, Israel
    \and Department of Geophysics, Raymond and Beverly Sackler Faculty of Exact Sciences, Tel Aviv University, Tel Aviv 6997801, Israel  
    \and Institut f\"ur Astrophysik, Georg-August-Universit\"at,  Friedrich-Hund-Platz 1, 37077 G\"ottingen, Germany 
    \and Landessternwarte, Zentrum f\"ur Astronomie der Universt\"at Heidelberg, K\"onigstuhl 12, D-69117 Heidelberg, Germany
    \and School of Physics and Astronomy, Raymond and Beverly Sackler Faculty of Exact Sciences, Tel Aviv University, Tel Aviv, 6997801, Israel
}

\date{Received 13 September 2019 / Accepted 16 January 2020}
 
\abstract
    {The High Accuracy Radial velocity Planet Searcher (HARPS) spectrograph is mounted since 2003 at the ESO 3.6\,m telescope in La Silla and provides {\it state-of-the-art} stellar radial velocity (RV) measurements with a precision down to $\sim1$\,m\,s$^{-1}$. The spectra are extracted with a dedicated data-reduction software (DRS) and the RVs are computed by cross correlating with a numerical mask.\looseness=-5}
    {The aim of this study is three-fold: 
(i) Create an easy access to the public HARPS RV data set.
(ii) Apply the new public SpEctrum Radial Velocity AnaLyser (SERVAL) pipeline to the spectra, and produce a more precise RV data set.
(iii) Check whether the precision of the RVs can be further improved by correcting for small nightly systematic effects. 
    }
    {For each star observed with HARPS, we downloaded the publicly available spectra from the 
    ESO archive, and recomputed the RVs with SERVAL. This was based on fitting each observed spectrum with a high signal-to-noise ratio template created by co-adding all the available spectra of that star. We then 
    computed nightly zero points (NZPs) by averaging the RVs of quiet stars.
    }
    {
    Analysing the RVs of the most RV-quiet stars, whose RV scatter is $<5$\,m\,s$^{-1}$, we find that SERVAL RVs are on average more precise than DRS RVs by a few percent. Investigating the NZP time series, we find three significant systematic effects, whose magnitude is independent of the software used for the RV derivation: (i) stochastic variations with a magnitude of $\sim1$\,m\,s$^{-1}$; (ii) long-term variations, with a magnitude of $\sim1$\,m\,s$^{-1}$ and a typical timescale of a few weeks; and (iii) $20$--$30$ NZPs significantly deviating by few m\,s$^{-1}$. In addition, we find small ($\lesssim1$\,m\,s$^{-1}$) but significant intra-night drifts in DRS RVs before the $2015$ intervention, and in SERVAL RVs after it.
    We confirm that the fibre exchange in $2015$ caused a discontinuous RV jump,
    which strongly depends on the spectral type of the observed star: from $\sim 14$\,m\,s$^{-1}$ for late F-type stars, to $\sim -3$\,m\,s$^{-1}$ for M dwarfs. The combined effect of extracting the RVs with SERVAL and correcting them for the systematics we find is an improved average RV precision: $\sim5\%$ improvement for spectra taken before the $2015$ intervention, and $\sim15\%$ improvement for spectra taken after it. To demonstrate the quality of the new RV data set, we present an updated orbital solution of the GJ\,253 two-planet system.
    }
    {Our NZP-corrected SERVAL RVs can be retrieved from a user-friendly, public database. It provides more than 212\,000 RVs for about 3000 stars along with many auxiliary information, such as the NZP corrections, various activity indices, and DRS-CCF products.
    }



\maketitle


\section{Introduction}
 
The High Accuracy Radial velocity Planet Searcher \citep[HARPS,][]{Pepe2002,Mayor2003Msngr.114...20M} operates since 2003 at the 3.6\,m telescope of the European Southern Observatory (ESO) in La Silla. It is the first fibre-fed high-resolution echelle spectrograph capable of measuring stellar radial velocity (RV) with a precision down to $\sim1$\,m\,s$^{-1}$. 
In this context, HARPS discovered  a plethora of exoplanets in the past 15\,years. 
More notably, HARPS proved to be an effective hunter of small and some even potentially temperate exoplanet systems, e.g. around GJ\,581 \citep{Bonfils2005, Udry2007}, GJ\,536 \citep{Mascareno2017}, Proxima Centauri \citep{Anglada2016Natur.536..437A}, and HD\,10180 \citep{Lovis2011}. 
With its unprecedented precision, HARPS is the southern hemisphere backbone Doppler validation instrument for the {\it Transiting Exoplanet Survey Satellite} \citep[{\it TESS};][]{Ricker2015}, which is uncovering hundreds of small rocky transiting exoplanet candidates around nearby stars.
HARPS also offers a large publicly available spectral archive, which has already allowed the post-detection Doppler validation of {\it TESS} candidates, for example  GJ\,143, HD\,23472 \citep[][]{Trifonov2019a}, HD\,15337 \citep{Gandolfi2019, Dumusque2019} and GJ\,357 \citep{Luque2019}.


The HARPS spectrograph is precise and stable on years-long timescales thanks to active environmental control (mainly temperature and pressure) and the stability of its atomic standard calibration, typically a ThAr hollow cathode lamp, and since 2011 a Fabry-Perot etalon \citep{Wildi2010SPIE.7735E..4XW}.
Although a laser frequency comb is available since April 2015 \citep{LoCurto2012} it is not in routine operation yet due to limited spectral coverage and fibre lifetime.

Despite HARPS' stability, its wavelength calibration, drift measurements, and cross-calibration, are nontrivial procedures and are a potential bottleneck limiting the RV precision. The shortcomings in the pipeline, the instrument, or the observations, may lead to systematic errors. An example is the so-called "CCD stitching" systematic, which was discovered with the laser frequency comb \citep{Wilken2010MNRAS.405L..16W} and which is not handled by the current pipeline and leads to $\sim$1\,yr periodicity in the HARPS RVs correlated with the barycentric Earth radial velocity \citep{Bauer2015A&A...581A.117B, Dumusque2015ApJ...808..171D,Coffinet2019}. Another well-known systematic was introduced with the upgrade of the optical fibres in May 2015 \citep{LoCurto2015}. This upgrade changed the instrumental profile, and thus the RV offset between the pre- and post-upgrade RVs. This offset is not the same for all stars and might depend on the stellar spectral type.
 
In this work, we take advantage of the large set of HARPS publicly available wavelength-calibrated 
spectra accumulated over the years by many programs and groups with different observing strategies and goals.
We analyzed the HARPS sample for common RV systematics in an attempt to deliver more precise Doppler measurements for the exoplanet community, and made these data available in a user friendly catalog.

We re-derived more than 212\,000 RVs from publicly available stellar HARPS spectra with the 
SpEctrum Radial Velocity AnaLyser \citep[SERVAL,][]{Zechmeister2018} pipeline.
We then looked for the presence of systematic effects by investigating HARPS' nightly zero point RVs (NZPs): a methodology we had established for CARMENES data \citep{Trifonov2018A&A...612A..49R} and for archival HIRES RVs \citep{Tal-Or2019MNRAS.484L...8T}.
Similar approaches had also been applied for the SOPHIE spectrograph  \citep{Courcol2015A&A...581A..38C, Hobson2018A&A...618A.103H}. The systematics revealed in those works can often be traced back and attributed to known instrumental events such as detector changes or fibre coupling problems.

In \autoref{Sec2} we introduce the publicly available HARPS data and the stellar sample we use in our analysis. In \autoref{Sec3} we introduce our HARPS spectral re-processing scheme with the SERVAL pipeline. In \autoref{Sec4} we present our findings regarding HARPS' systematic RV variations. In \autoref{Sec5} we present the main results of our work, and \autoref{Sec6} gives a brief summary and draws some conclusions.

\begin{figure}[ht]
    \centering
    \includegraphics[width=9cm,height=7.5cm]{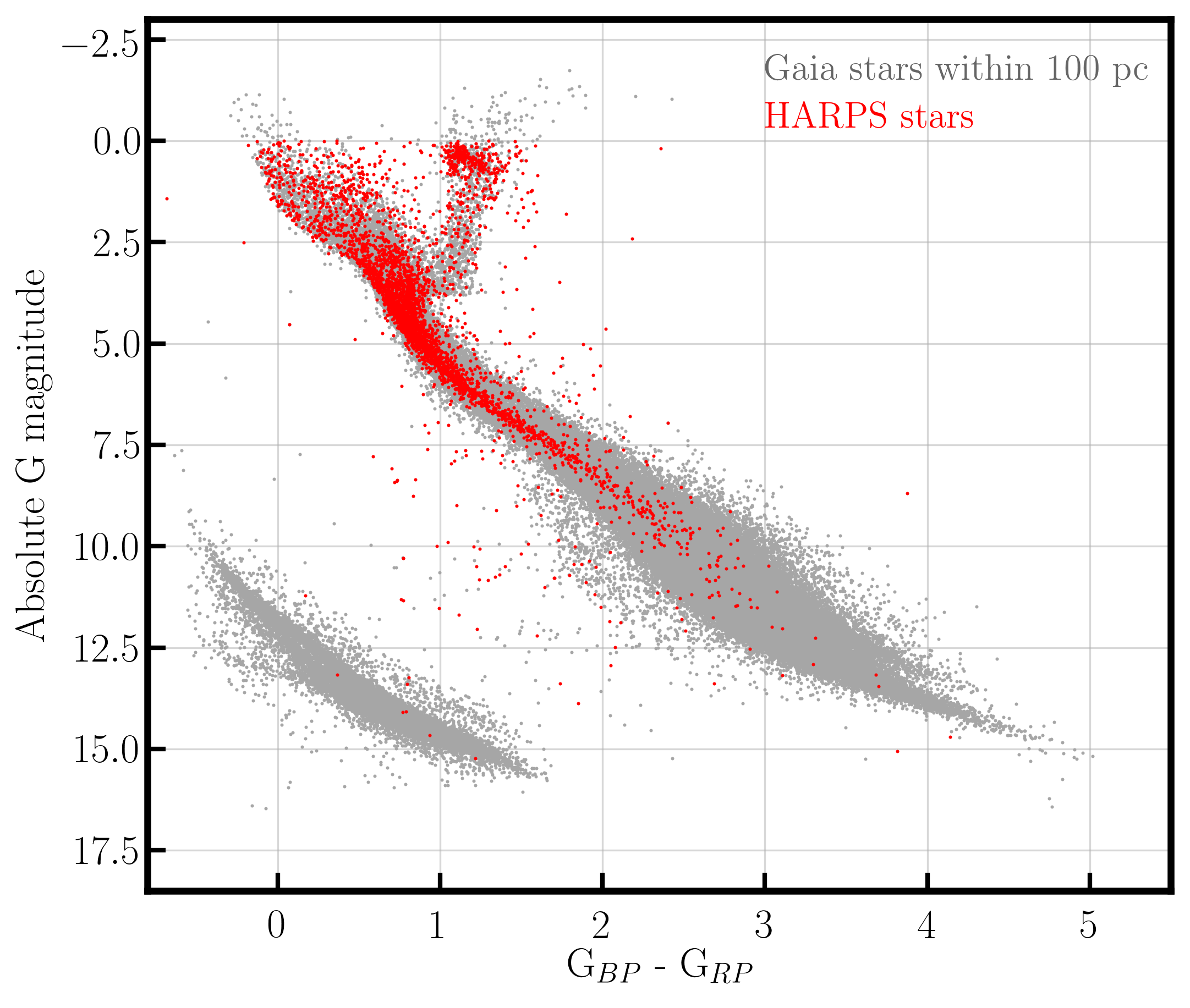}   
    \caption{Hertzsprung-Russel diagram of 212\,728 stars within 100\,pc observed by {\em Gaia} (grey dots). The red dots represent 5260 stars observed with HARPS, which we identified in the 
    {\em Gaia-DR2} catalog.
    }
    \label{fig:HARPS_HR}
\end{figure}

\begin{figure*}[ht]
    \centering
    \includegraphics[width=0.33\linewidth]{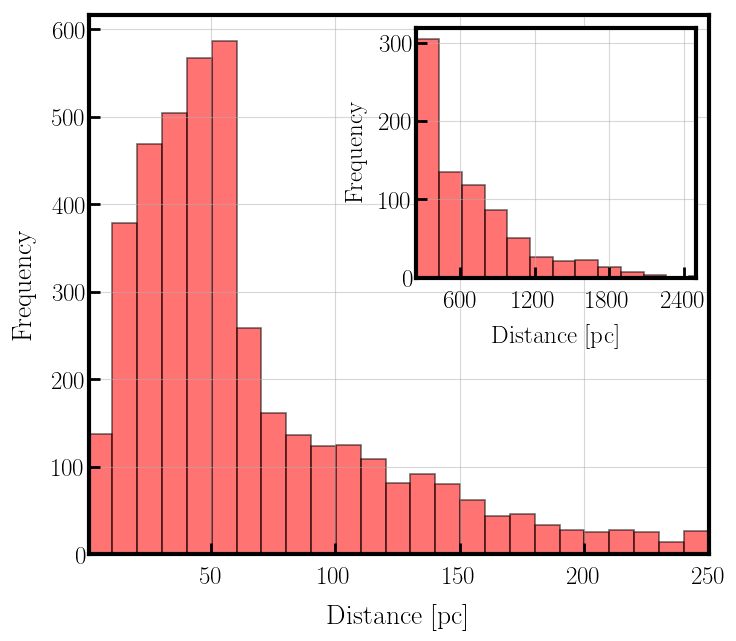}    
    \includegraphics[width=0.33\linewidth]{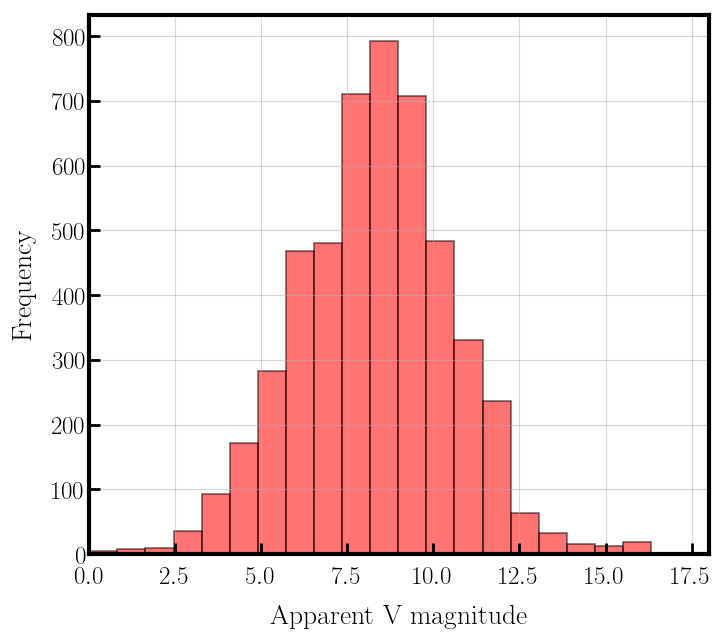}
    \includegraphics[width=0.33\linewidth]{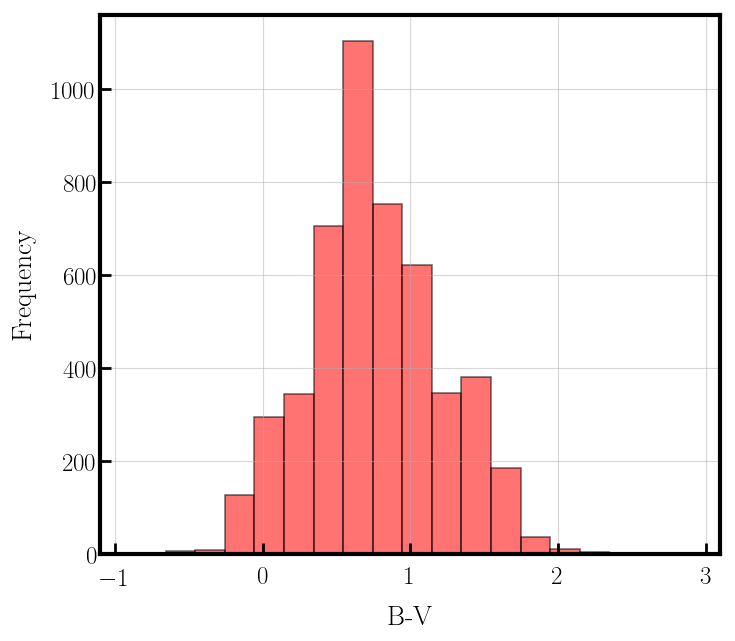}
    \caption{Properties of the HARPS sample of 5260 stars used in this work. The panels shows the distribution of the {\em Gaia} distances (\emph{left}), of the apparent $V$ magnitudes (\emph{middle}), and of the $B-V$ color (\emph{right}).
    }
    \label{fig:HARPS_sample}
\end{figure*}

\section{The HARPS data and the stellar sample}
\label{Sec2}

Once public, HARPS spectra can be queried and downloaded from the ESO archive using the generic query form\footnote{\url{http://archive.eso.org/wdb/wdb/adp/phase3_main/form}}, which provides access to all phase 3 data reduced by HARPS' data-reduction software (DRS v3.5). The data products also contain detailed auxiliary information regarding the observation such as target's coordinates, Earth's barycentric radial velocity, signal-to-noise ratio (S/N) in each order, drift measure, etc.
Furthermore, the DRS provides a precise RV measurement derived by the spectrum cross-correlation function (CCF) method using a weighted binary mask \citep{Pepe2002}, as well as the CCF's full-width half-maximum (FWHM) and the Bisector Inverse Slope span (BIS-span) measurements. The FWHM and BIS are often used as stellar activity indicators \citep[e.g.][]{Queloz2001}. 


We have developed a new pipeline, which downloads, extracts, and re-processes the 
large public HARPS archive in a consistent way, and creates an easy user access to its scientific products such as high-precision Doppler measurements and activity indices  (see \autoref{HARPS_database}).
From the ESO-HARPS archive, we have downloaded a total of $264\,058$ publicly available reduced two-dimensional, multi-order spectra of more than $6\,100$ objects, which were observed with HARPS between 2003 and mid-2018. 
We have excluded from our analysis spectra which were not useful for high-precision stellar 
RV statistical analysis, such as solar spectra, asteroids, the Galilean moons, quasars and supernova candidates, which had been obtained in different astrophysical contexts.
Overall, we identified 5260 stellar targets in the HARPS sample.
Additionally, we excluded stars with less than three usable spectra -- the minimum requirement for creating a meaningful spectral template and thereafter precise RVs with SERVAL. Eventually, we have selected a total of $\sim3\,000$ reliable targets of F, G, K, M and L spectral types, to use for our subsequent NZP analysis.

Figure~\ref{fig:HARPS_HR} shows a  Hertzsprung-Russel (HR) diagram of the HARPS sample stars over-plotted on top of the known stars within 100\,pc, retrieved from the {\em Gaia DR2} catalog \citep{Gaia_Collaboration2016, Gaia_Collaboration2018b}.
During its operational time, HARPS observed mainly bright nearby main-sequence stars and late-type G and K giant stars. Other stellar objects, such as white dwarfs, very faint late-type M-dwarfs, and brown dwarfs are less suitable for a 3.6 m class telescope and HARPS.

Figure~\ref{fig:HARPS_sample} describes the HARPS stellar sample in terms of distributions of the estimated stellar distances \citep{Bailer-Jones2018}, $V$ magnitudes and $B - V$ colors. 
It is evident that the HARPS surveys conducted in the past 15 years focused mainly on nearby main-sequence stars of spectral types G0\,V to M6\,V, with a median $B - V$ color of 0.722\,mag, median apparent V magnitude of 8.4 \,mag, and with a median distance of $\sim$120 pc. 
The stars within 100 pc represent $\sim$ 67\%  of our HARPS sample.
This collection of stars, are representative of a volume-limited, long-lasting HARPS surveys dedicated to solar-mass G-dwarf stars \citep{Pepe2004,LoCurto2010,Moutou2011,LoCurto2013,Udry2019}, 
low-mass M-dwarfs \citep{Bonfils2005,Mayor2009,Forveille2009,Bonfils2013, Anglada2016Natur.536..437A, Astudillo-Defru2017, Ribas2018} and 
metal-poor stars \citep{Santos2011,Faria2016,Mortier2016}, which all target nearby stars.
The remaining $\sim$ 28\% of the HARPS sample (with distance $>$ 120\,pc) are typically  
bright main sequence stars of spectral types A0 to F6, some fainter and more distant transiting planet hosts observed by more recent RV follow-up campaigns of transit planet candidates from the HATSouth \citep{Bakos2013,Brahm2016,Henning2018,Espinoza2019}, WASP-south \citep{Pollacco2006,Gillon2009,Nielsen2019}, and the {\sc K2} extended mission \citep{Howell2014, Grziwa2016, Johnson2018}, or evolved sub-giant and giant branch stars of spectral types G8\,IV $-$ K4\,III.

\begin{figure*}[ht]
    \centering
    \includegraphics[width=6cm,height=5cm]{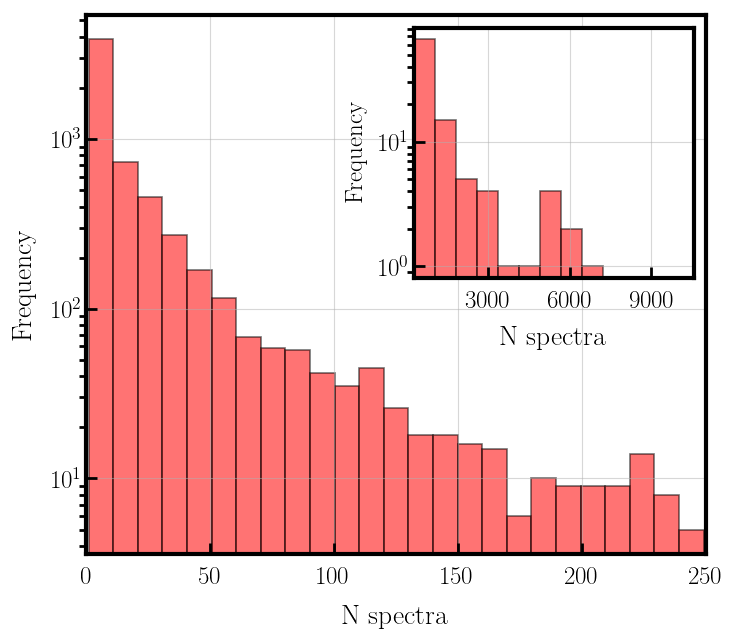}
    \includegraphics[width=6cm,height=5cm]{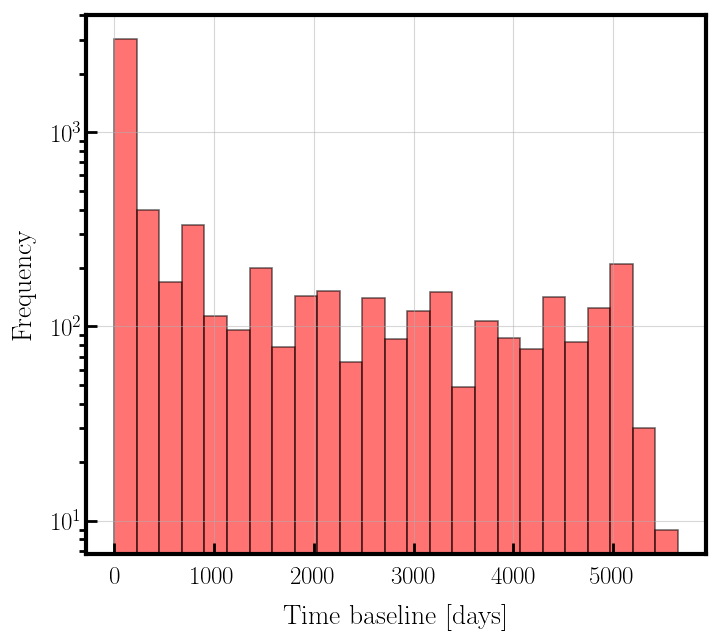}
    \includegraphics[width=6cm,height=5cm]{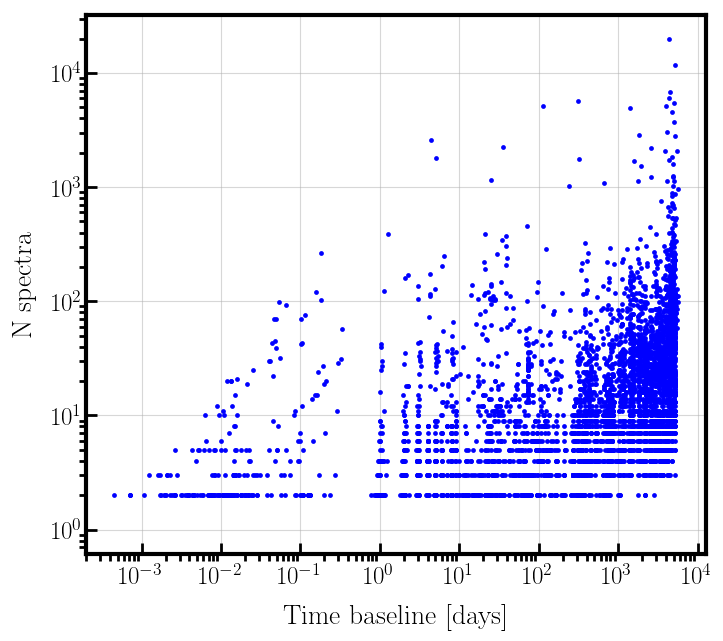} 
    \caption{Observation statistics of the HARPS target sample of 5260 stars used in this work. The panels show the distribution of the number of spectra per target $(left)$, the time baselines of the observed targets $(middle)$, and a scatter plot  of the time baseline against the number of spectra per target $(right)$.
    }
    \label{fig:HARPS_nobs}
\end{figure*}

\begin{figure*}[ht]
    \centering
    \includegraphics[width=6cm,height=5cm]{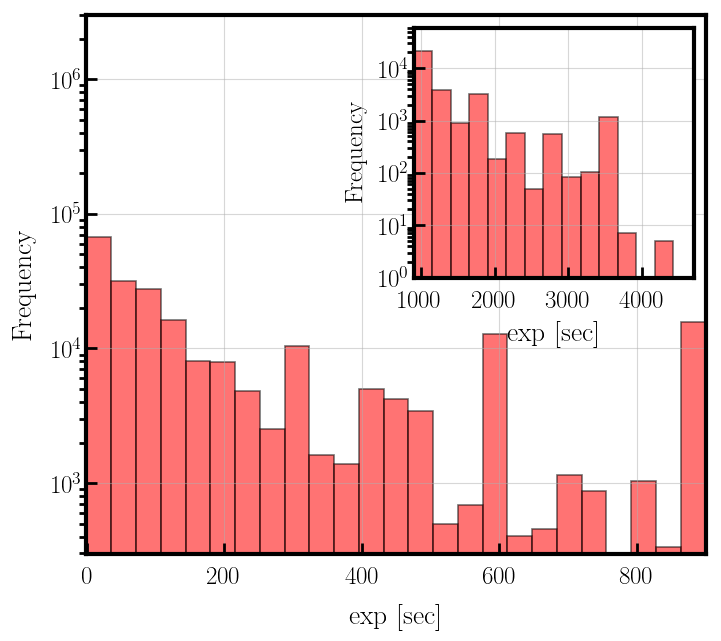}
    \includegraphics[width=6cm,height=5cm]{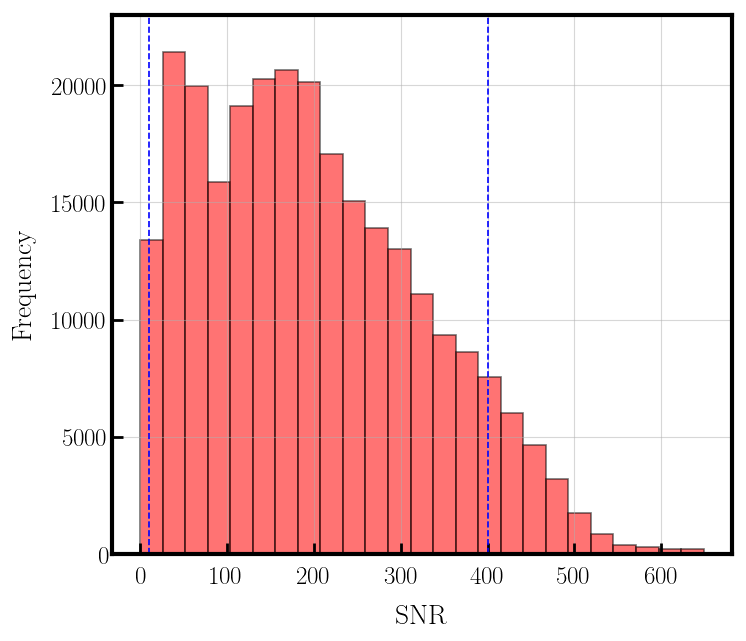}
    \includegraphics[width=6cm,height=5cm]{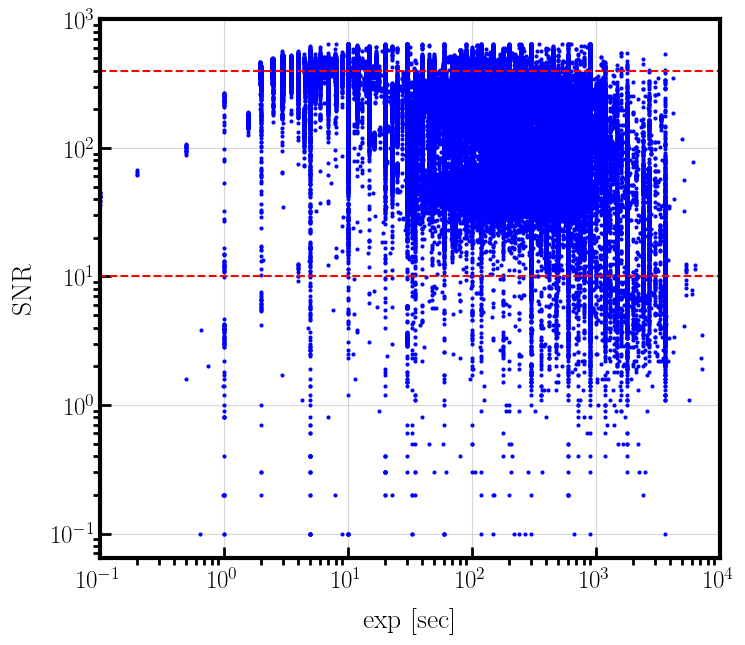} 
    \caption{Observation statistics of the HARPS spectra used in this work ($\sim$ 212\,000 spectra). The panels show the distribution of the exposure times $(left)$, the achieved S/N at 550 nm for the obtained spectra $(middle)$, and a scatter plot of the exposure times versus the S/N $(right)$. The red dashed lines mark the range of S/N adopted in this work to create a template spectrum. 
    }
    \label{fig:HARPS_nobs2}
\end{figure*}

Figure~\ref{fig:HARPS_nobs} shows some basic observational statistics from the employed HARPS sample.
The left panel shows a histogram of the number of spectra  per target, whereas  
the middle panel shows a histogram of the time baselines of the observed targets.
The scatter plot in the rightmost panel shows the relation between the two quantities.
It is evident from the Figure that the majority of the targets observed in the sample have fewer than 50 spectra, with about 38\% of the targets having fewer than 10 spectra. 
Some of the targets have very short
time baseline below 2 days, but nevertheless often have a sufficient number of spectra.
These are likely the result of special observational campaigns such as transit spectroscopy 
or observations of the Rossiter-McLaughlin \citep[RM;][]{Rossiter1924,McLaughlin1924,Queloz2000} 
effect of known transiting planets, which require
numerous consecutive observations within one or two nights.
Most of the stars have time baseline longer than three years and those 
also tend to have many spectra. The majority of these are most likely 
known single or multiple-planet hosts and "RV-standard" stars, which show a very low RV scatter, 
and thus are frequently observed  for instrument stability monitoring over the past 15 years of HARPS service.
For example, the most frequently observed target with a total 
of $19\,640$ spectra that stood out on the top right corner in the right panel of Fig.~\ref{fig:HARPS_nobs} is our close neighbour $\alpha$ Cen B \citep{Dumusque2012,Rajpaul2016}.
The second most observed star with a total of $11\,666$ spectra is the RV-standard $\tau$\,Ceti (HD\,10700), which according to \citet[][]{Tuomi2013} has up to five Super-Earth planets.
The list of frequently observed stars is followed by another RV standard --
$\epsilon$\,Eri (HD\,20794) -- with a total of $6\,775$ spectra,
which is also a potential multi-planet systems \citep{Pepe2011,Feng2017},
followed by $\delta$ Pav (GJ\,780) with $6\,003$ spectra, 
and many other long-term observed planetary systems \citep[e.g.][]{Udry2019}.

Figure~\ref{fig:HARPS_nobs2} shows statistics of the exposure times and S/N at 550\,nm of the public HARPS spectra. The left panel of 
Fig.~\ref{fig:HARPS_nobs2} shows that in addition to the most commonly used exposure times of 300, 600, 900, 1200 and 1800 seconds, there is an 
abundance of exposures with less than 150 seconds. 
Again, these are likely spectra of the most 
heavily observed RV standard stars, which are very bright, and hence the short 
exposures and large number of observations.
The middle panel of Fig.~\ref{fig:HARPS_nobs2} shows a histogram of the typical S/N at 550\,nm calculated by the DRS pipeline. 
The distribution seems to be bimodal with a peak near S/N $\sim$ 50, and another broader peak near S/N $\sim$ 180. 
Overall, the stars in our HARPS sample are bright (see middle panel in \autoref{fig:HARPS_sample}), 
and the exposure times were selected for achieving the high S/N needed for maximum RV precision. 
The right panel of \autoref{fig:HARPS_nobs2} shows a scatter plot 
of the exposure times and the S/N. The red-dashed lines mark 
the range of 10 $<$ S/N $<$ 400 adopted in our analysis for 
the creation of the stellar template needed to derive precise RVs with SERVAL (see Sect.~\ref{Sec3}).

\section{Deriving RVs and activity indicators with SERVAL}
\label{Sec3}

In addition to HARPS DRS RVs, we also derived precise RV measurements with SERVAL \citep[][]{Zechmeister2018} based on the same DRS spectrum extraction and the same wavelength solution. Instead of using a pre-calculated numeric mask, SERVAL creates for each observed star a high S/N template spectrum by shifting and co-adding all individual spectra of that star. The template is then used to derive the RVs from the same observed spectra by using a $\chi^2$-minimization approach.
SERVAL is a data-driven approach that aims to exploit all the RV information in a self-consistent way. It post-processes the data, requires at least a few spectra, and provides differential RVs. In contrast, the DRS provides absolute RVs with an excellent precision in an online fashion and given a proper choice of a mask and an initial RV guess. 
Yet, \citet{Escude2012} demonstrated that the co-adding method can provide higher RV precision than the CCF method with a weighted binary mask employed in the standard HARPS DRS pipeline, in particular for late-spectral-type stars.
We compare the DRS and SERVAL RVs for a subset of the most quiet and heavily observed stars in our sample in \autoref{HARPS_database}.

Due to the significant change in HARPS' instrumental profile that accompanied the fibre upgrade in 2015 \citep{LoCurto2015}, we applied SERVAL separately to spectra obtained before and after the intervention. Hence, for each star that was observed both before and after the intervention, we created two high  S/N templates.

For the template creation and the NZP analysis we use only spectra within the range 10 $<$ S/N $<$ 400. 
Here we aim to co-add only high-quality spectra and avoid biases from noisy or saturated spectra.
We do derive RVs from observations within the range 3 $<$ S/N $<$ 10 and 400 $<$ S/N $<$ 500, but these were flagged as low-S/N and high-S/N observations, respectively, and must be taken with caution. Spectra with S/N $<$ 3 and S/N $>$  500 were not considered for SERVAL analysis.

\begin{figure*}
    \centering
    \includegraphics[width=14cm,height=7cm]{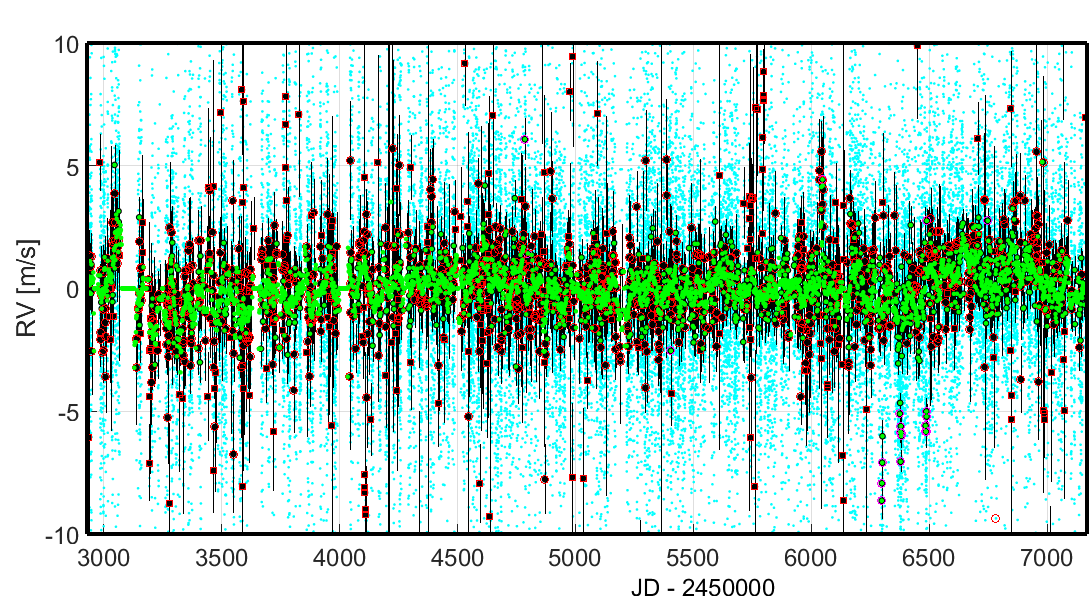}\includegraphics[width=3.5cm,height=7cm]{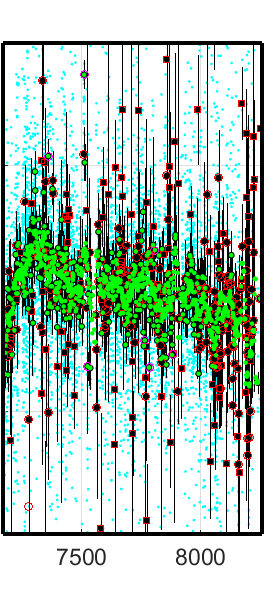}
    \newline
    \includegraphics[width=14cm,height=7cm]{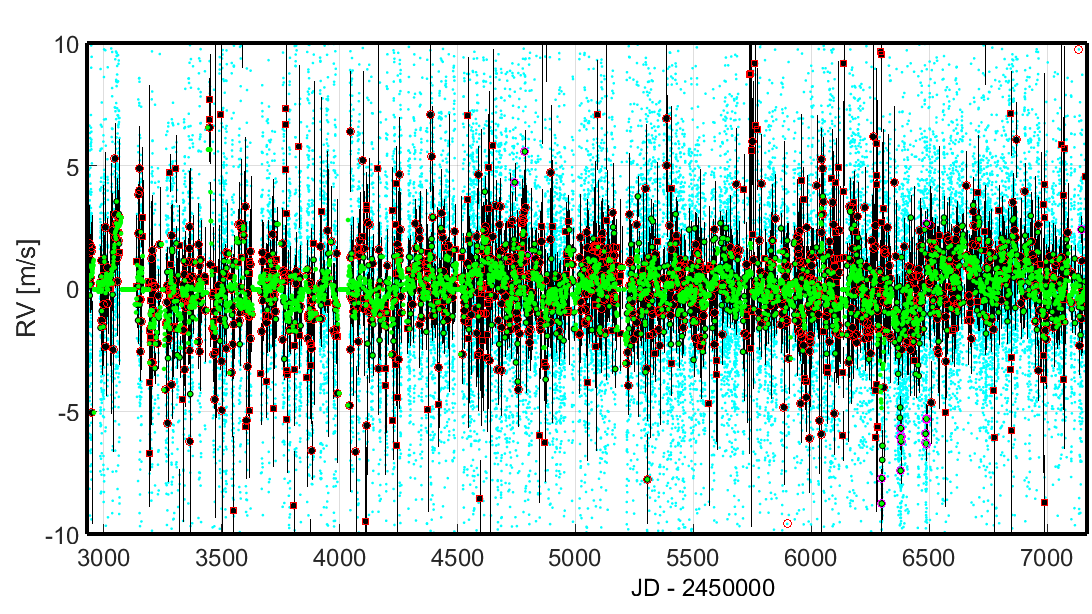}\includegraphics[width=3.5cm,height=7cm]{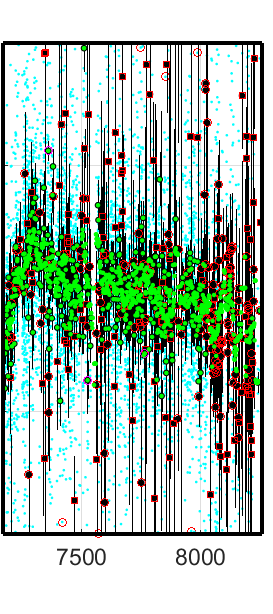}
    \caption{HARPS NZPs based on SERVAL RVs (\emph{upper panel}) and DRS RVs (\emph{lower panel}). The stellar zero-point subtracted RVs (cyan dots) of all RV-quiet stars have been averaged in each night (NZPs, black points). NZPs calculated with too few RVs ($n_{{\rm RV,}n}<3$, red boxes) or NZPs with too large uncertainties ($\gtrsim1$\,m\,s$^{-1}$, red circles) are excluded from correcting the stellar RVs. The RVs in these red-marked nights are corrected by using a smoothed version of the NZP curve, which was calculated with a moving weighted-average ($21$\,d window). For significantly deviating NZPs (magenta circles) we adopted their individual (unsmoothed) NZP value, regardless of their uncertainties. The green dots denote for each night the eventual NZP value that was used for the stellar RV correction. Note the broken x-axis at JD\,$\sim2\,457\,169$ (May 26, 2015): since HARPS-pre and HARPS-post NZPs were calculated separately, which also included subtracting each star's zero-point RV, both time series are centered around\,$\sim0$\,m\,s$^{-1}$, despite the jump in stellar absolute RVs (see \autoref{sec:RVjump}). The RV axis was limited to $\pm10$\,m\,s$^{-1}$ to highlight the small NZP variations.
    }
    \label{fig:HARPS_serval_nzp}
\end{figure*}

The SERVAL pipeline also measures stellar activity indicators, which serve as important 
diagnostics of the planetary induced Doppler signal hypothesis.
For the HARPS spectra SERVAL provides a measure of emission in the H$\alpha$, 
Na\,I D, and Na\,II D activity-related lines. To measure these activity indicators, SERVAL needs an estimate of the stellar absolute RV which is requested from the fits header (here DRS-CCF) or as a fallback from {\sc SIMBAD}. In addition, SERVAL measures the differential line width (dLW), quantifying variations in the spectral line widths, and the chromatic RV index (CRX) of the spectra, which provides an information about the wavelength dependence of the RV from individual spectral orders as induced by e.g. spots.
All the spectral diagnostics obtained with SERVAL come with their uncertainties.
For more detailed description of the SERVAL activity indicators we refer to \citet{Zechmeister2018}.

\section{Systematic effects in HARPS RVs}
\label{Sec4}

\subsection{\label{subsec:NZPmethod}Nightly zero-point variations}

Since the calibrations for the HARPS observations are typically
done at the beginning of each night, we focused our search for systematic effects by calculating the time series of nightly zero-point RVs (NZPs).
To calculate the NZPs we followed a similar procedure to the one described in \citet{Tal-Or2019MNRAS.484L...8T}, where a more detailed description of the algorithm can be found. In short, we calculated a NZP for each night in which at least three different RV-quiet stars (RV scatter $<10$\,m\,s$^{-1}$) were observed, by taking the weighted average of the RV, after subtracting from each star its own weighted-average RV (stellar zero point).

Unlike the case of HIRES data, where we had calculated only one NZP time series, for HARPS we had to calculate four different NZP time series: distinguishing between DRS-CCF and SERVAL RVs, and between RVs before the $2015$ intervention ("pre RVs") and after the intervention ("post RVs"). The latter separation was also done to prevent the stellar zero-point subtraction from influencing the NZP estimation. Since the intervention introduced spectral-type dependent discontinuous jumps to the stellar RV time series, the stellar zero point of each star in a time series that contains both pre and post RVs heavily depends on the fraction of observations taken before or after the intervention (see also \autoref{sec:RVjump}). Such inconsistent stellar zero points would add scatter to the NZP time series, which would enhance the NZP uncertainties. In addition, we calculated the four NZP time series for each RV-quiet star individually, after excluding the star itself from the NZP calculation process, to avoid self biasing.


Figure \ref{fig:HARPS_serval_nzp} shows, separately for SERVAL and DRS RVs, the RVs that were used to calculate the NZPs, the derived individual NZPs, and the NZPs that were actually used to correct the originally-derived RVs. The NZPs that were not used for RV correction are marked in red. These are either NZPs that were calculated with too few RV-quiet stars ($n_{RV,n}<3$) or NZPs with too large uncertainties ($\gtrsim1$\,m\,s$^{-1}$). The exact uncertainty threshold that we used to exclude a NZP from the correction stage slightly differs between the four different time series, since it was taken as the scatter of the NZPs around a smoothed version of their time series, which was calculated with a moving weighted-average ($21$\,d window) filter. For correcting the originally-derived RVs, we replaced the excluded NZPs with the filter's value in these nights, and fixed their NZP uncertainties to the scatter mentioned above. On top of that, about two dozen NZPs, significantly deviating from the filter, were adopted for the correction even if their uncertainties were above the threshold. Most of these outlier NZPs
come from three week-long observing runs performed in $2013$.

\begin{figure}
    \centering
    \includegraphics[width=1\linewidth,trim=0 28 54 0,clip]{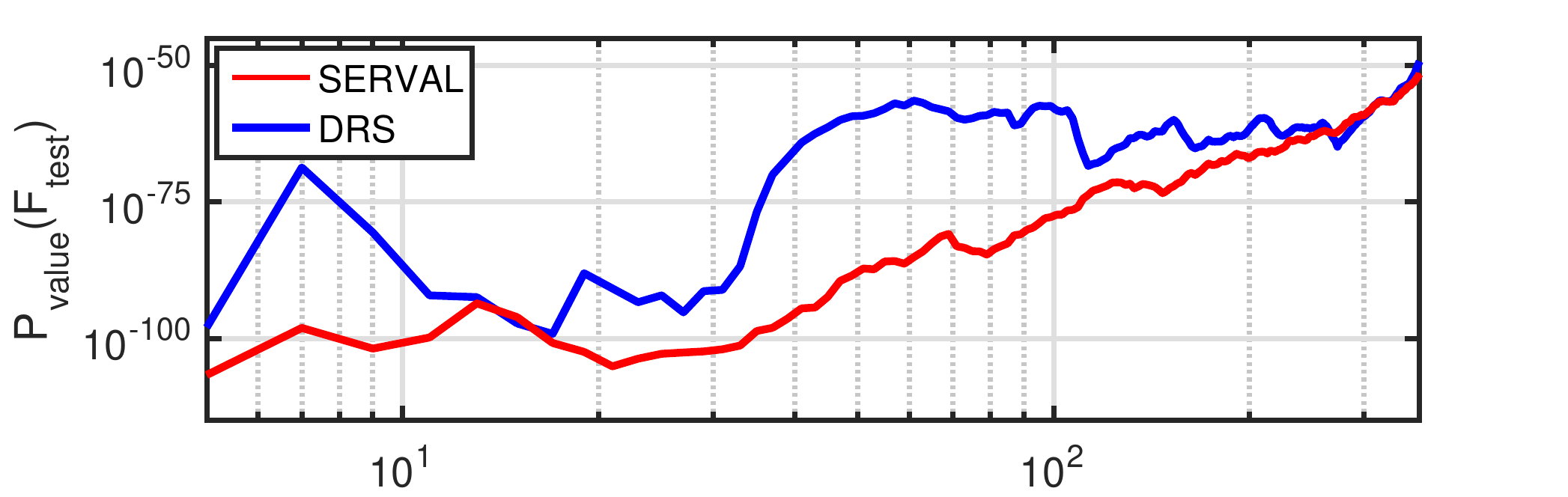}\\
    \includegraphics[width=1\linewidth,trim=0  0 54 0,clip]{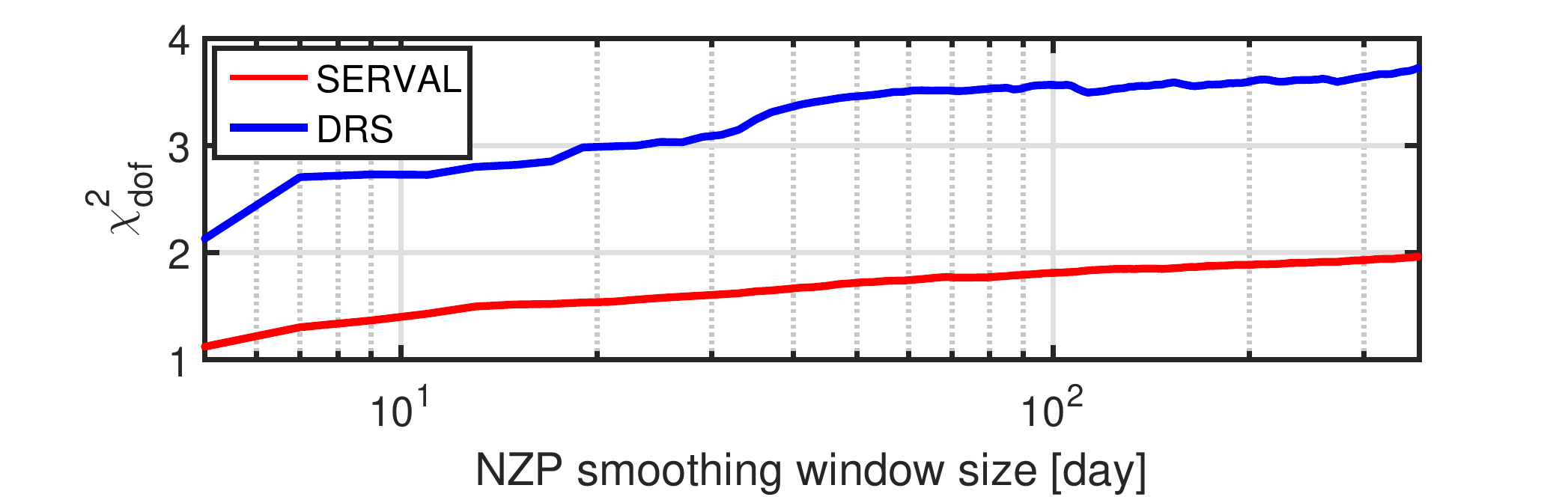}
    \caption{Optimizing the window size of the smoothing filter for HARPS-pre RVs.  \emph{Upper panel}: $p(F_{\rm test})$-value. \emph{Lower panel}: $\chi^2_{\rm dof}$. Both statistics were calculated for 'modeling' the NZP time series with a moving weighted-average filter.}
    \label{fig:HARPS_win_optim}
\end{figure}

This somewhat elaborate NZP correction strategy is tailored to the apparent characteristics of HARPS' systematics. Similarly to \citet{Tal-Or2019MNRAS.484L...8T}, we searched for the typical timescale of HARPS' NZP variations by varying the window size of the filter from $3$ to $365$ days and looking at the $\chi^2_{\rm dof}$ and the $p(F_{\rm test})$-value statistics of subtracting the filter from the NZPs. A moving weighted-average filter can be viewed as a non-parametric model of the data, with an effective number of parameters equal to the data time-span divided by the window size. To illustrate the search, Fig.~\ref{fig:HARPS_win_optim} shows the results for HARPS-pre RVs. Unlike the case of the HIRES data, where both statistics indicated a similar typical timescale of $1$--$2$ month (see Fig. 3, \citet{Tal-Or2019MNRAS.484L...8T}), for the HARPS NZPs the two statistics are not entirely consistent with each other. While the $\chi^2_{\rm dof}$ statistic always preferred a smaller window, pointing towards a stochastic behaviour of the NZPs, the $p(F_{\rm test})$-value statistic showed a secondary minimum at a timescale of a few weeks, pointing towards a smooth NZP variation of the instrument. This behaviour probably indicates two effects that drive HARPS' NZP variations: a stochastic one, possibly related to calibration errors, and an additional effect related to more slowly varying instrumental drifts.

\begin{table}
    \caption{Main characteristics of the four NZP sets.}
    \label{table:nzp_stats}
    \centering
    \begin{tabular}{@{}l r r r r@{}}
        \topline 
                     &    \multicolumn{2}{c}{SERVAL}   &  \multicolumn{2}{c}{DRS}   \\
        Parameter    &    pre   & post  & pre  & post  \\
        \midline 
        &\multicolumn{4}{c}{Statistics}\\[.3ex]
        wrms(NZP) [m\,s$^{-1}$]                           &  1.32 & 1.58 & 1.44 & 1.59  \\
        med($\delta{\rm NZP}$) [m\,s$^{-1}$]              &  0.82 & 0.75 & 0.90 & 0.78 \\
        max($\delta{\rm NZP}$) [m\,s$^{-1}$]              &  0.87 & 0.99 & 1.00 & 1.09 \\[0.5ex]
        \midline 
        &\multicolumn{4}{c}{Night flags}\\[0.3ex]
        0     & 1352 &  426 & 1317 &  430 \\
        1     &   22 &    8 &   19 &    5 \\ 
        2     & 1018 &  182 &  911 &  168 \\
        3     & 1659 &  459 & 1753 &  458 \\
        4     &  176 &    4 &  227 &   18 \\
        total & 4227 & 1079 & 4227 & 1079 \\  
        \bottomline
    \end{tabular}
    \tablefoot{
       The meaning of the flags:\\
       0 - good NZP (small error and not an outlier);\\
       1 - outlier NZP;\\
       2 - too uncertain NZP, $\delta{\rm NZP}>\max(\delta{\rm NZP})$;\\
       3 - not enough RV quiet stars to calculate a NZP ($n_{RV,n}<3$); \\
       4 - no NZP or filter value were calculated (inside an observing gap $>21$\,d).
    }
\end{table}

The four NZP time series that were actually used to correct the originally-derived RVs are given in an online Table\footnote{\label{note1}\url{http://cdsarc.u-strasbg.fr/XXXXXXXXX}}. For each night since ${\rm JD} = 2\,452\,936$ (Oct. 23, 2003), the Table provides its NZP together with the uncertainty ($\delta{\rm NZP}$), the number of RV-quiet stars used to calculate the NZP ($n_{RV,n}$), and a flag specifying the type of the NZP. \autoref{table:nzp_stats} summarizes the main characteristics of the four NZP sets. Its upper panel shows the weighted rms (wrms), median NZP uncertainty (med($\delta{\rm NZP}$)), and the uncertainty threshold that was used to exclude a NZP from being used (max($\delta{\rm NZP}$)). The lower panel shows the number of nights with a certain flag. The five different flags are explained at the bottom of \autoref{table:nzp_stats}. The fact that the NZPs reveal a significant source of systematic RV scatter, on top of the internal uncertainties, is expressed in the ratio wrms(NZP)/med($\delta{\rm NZP}$), which is in the range of $1.6$--$2.1$ for the four RV sets.

\subsection{Average intra-night drift}

After correcting the RVs for NZP variations, as explained above, we checked the residual RVs for average intra-night drifts. A correlation between the RVs and the time relative to the local midnight $t_{\rm mid}$ is a good indicator not only of actual nightly drifts of the spectrograph, but also of seasonal effects, such as correlations of the RVs with the BERV or the hour angle of observation.

\begin{figure*}
    \centering
    \includegraphics[width=0.45\linewidth]{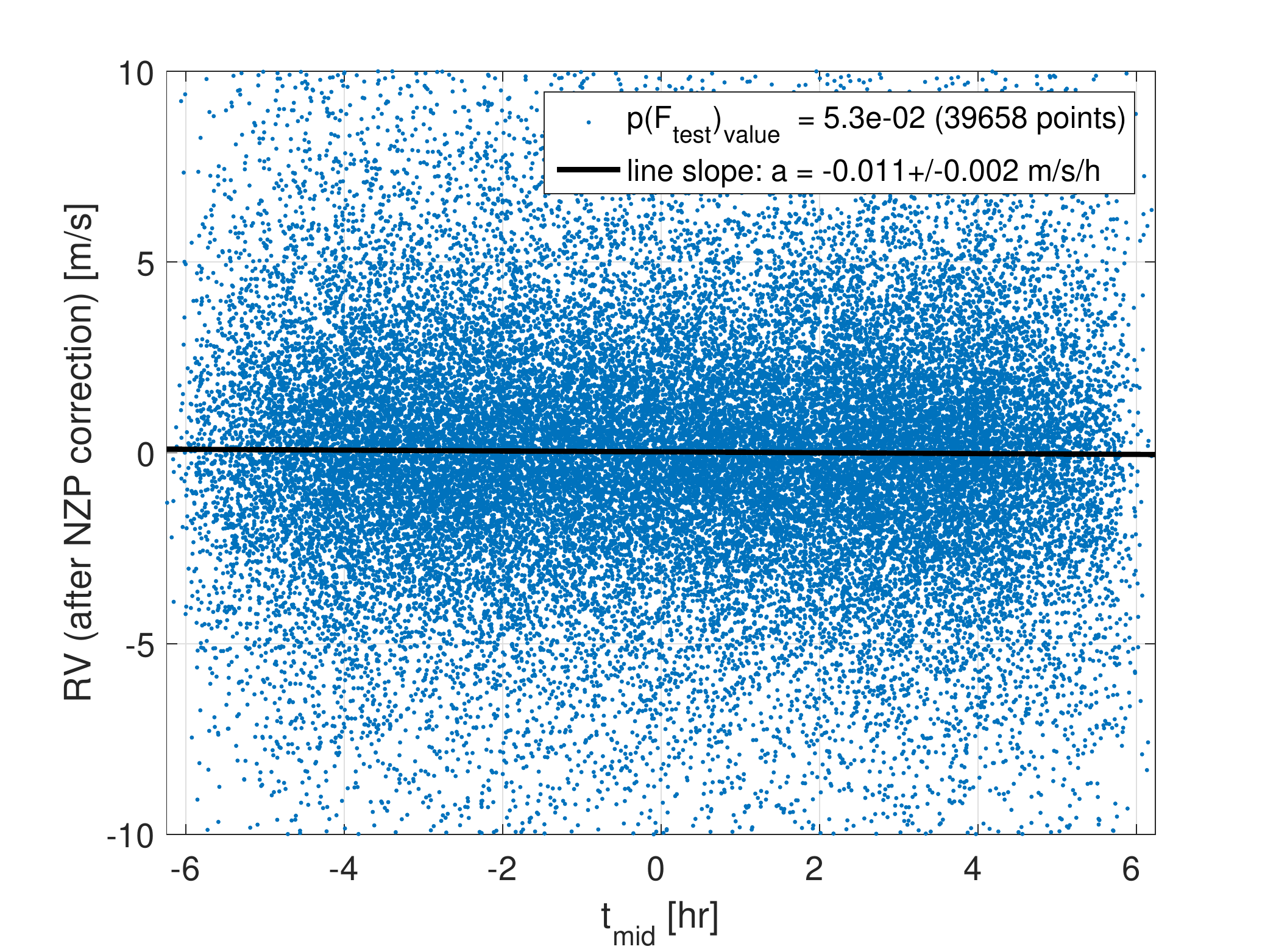}\includegraphics[width=0.45\linewidth]{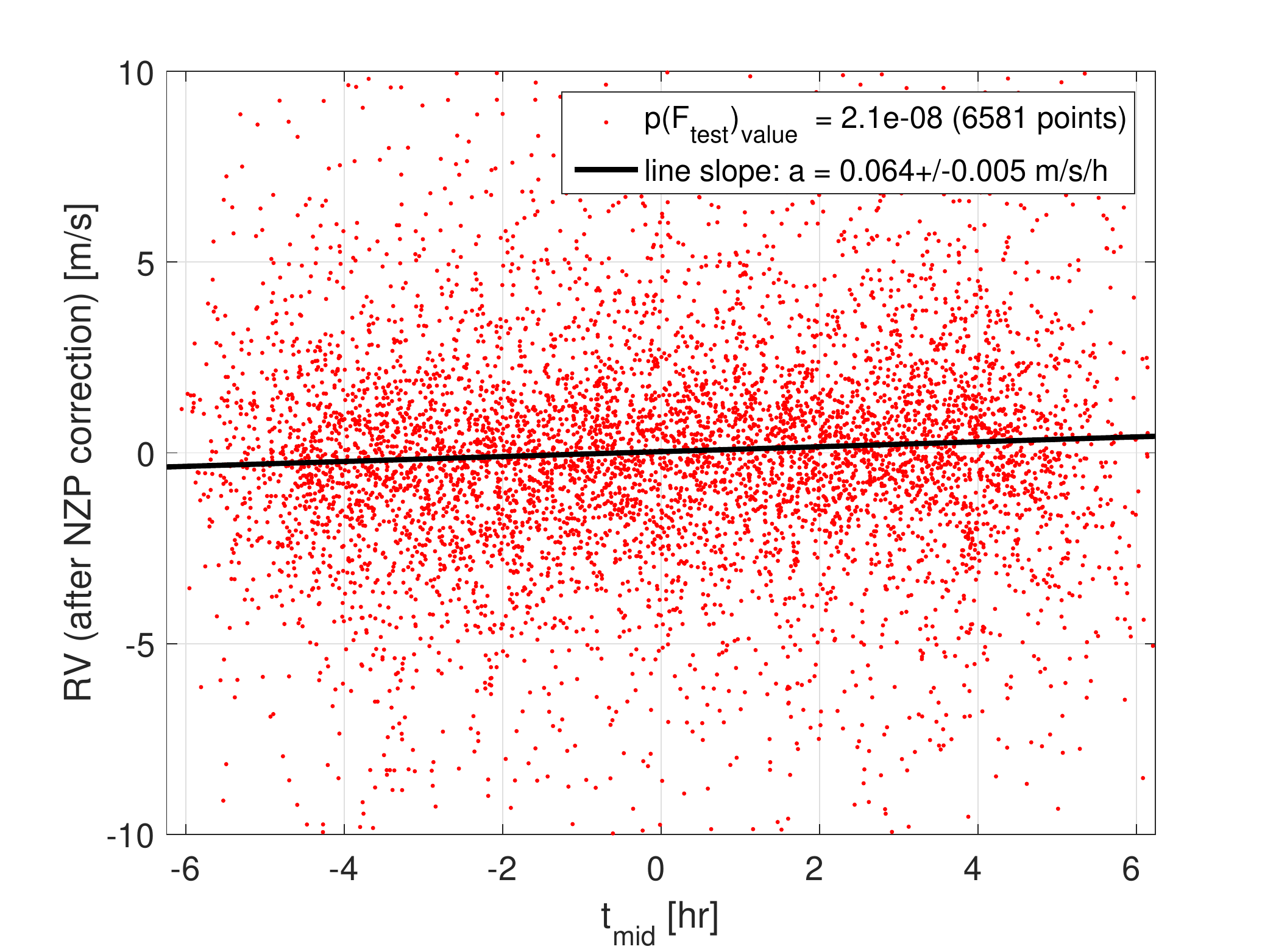}
    \caption{The average nightly drift in HARPS-SERVAL RVs. \emph{Left}: HARPS-pre. \emph{Right}: HARPS-post. Only RV-quiet stars (RV scatter $<10$\,m\,s$^{-1}$) were used for the line fits.}
    \label{fig:HARPS_serval_drift}
\end{figure*}

\begin{figure*}
    \centering
    \includegraphics[width=0.45\linewidth]{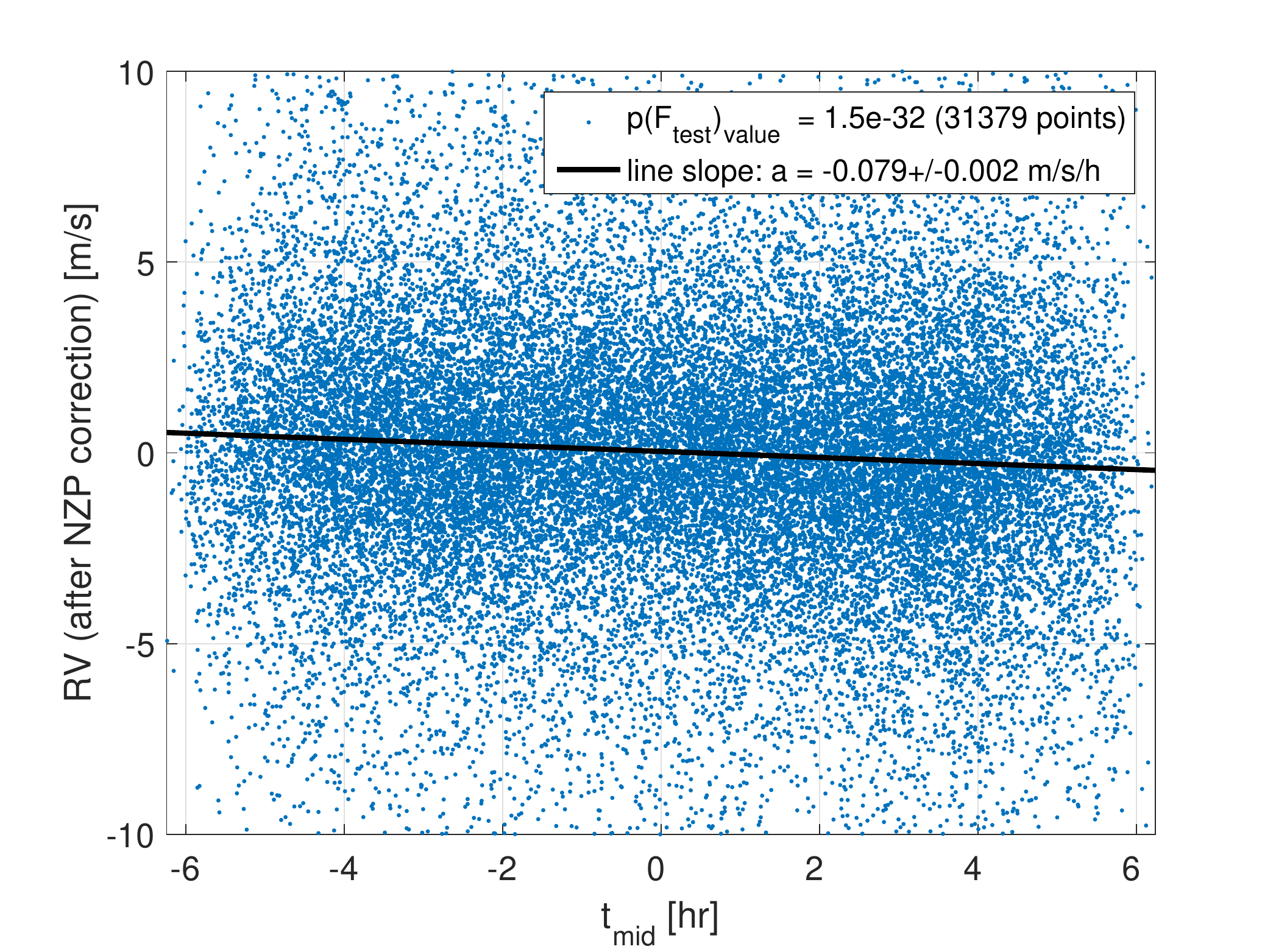}\includegraphics[width=0.45\linewidth]{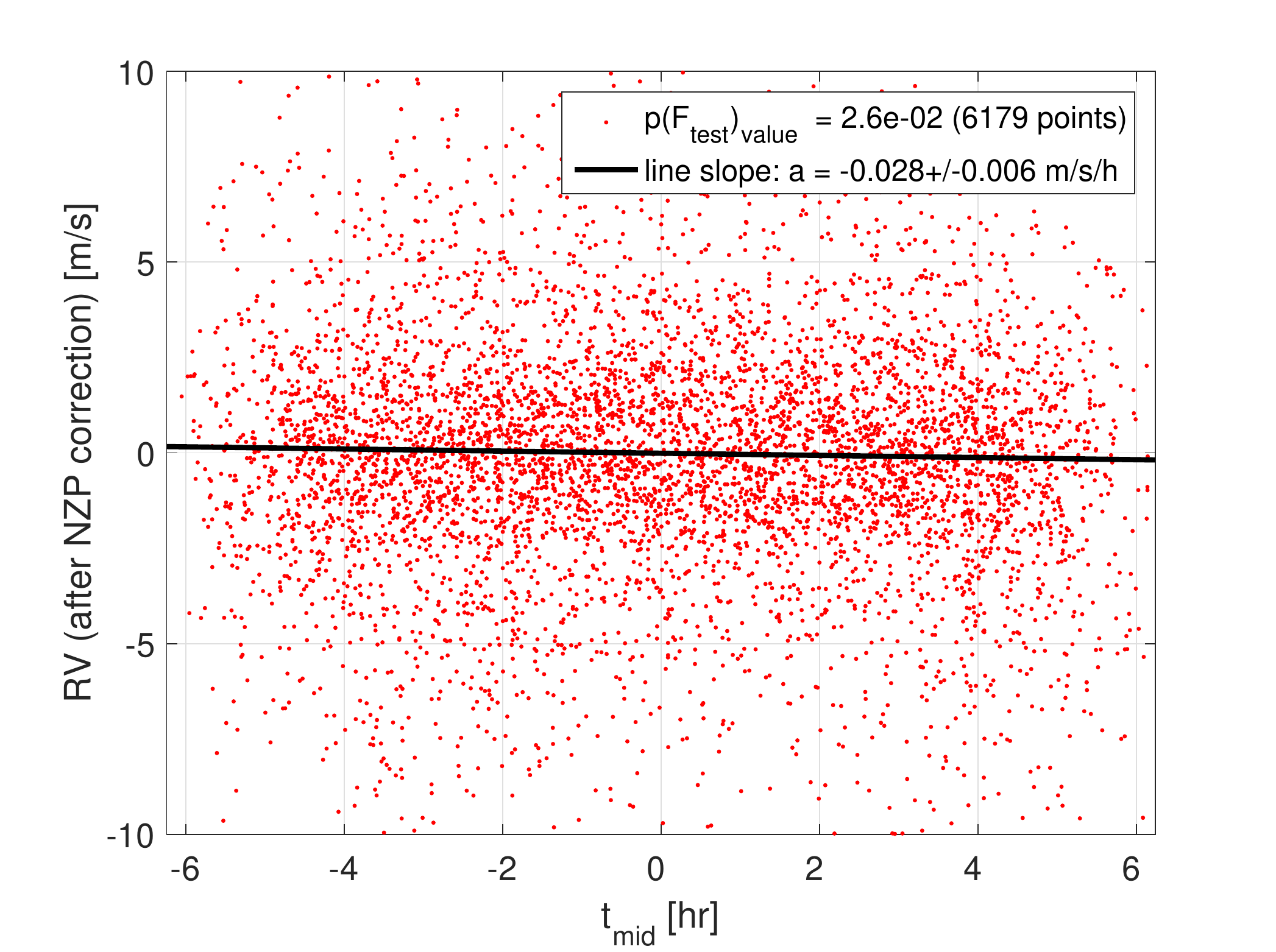}
    \caption{Same as \autoref{fig:HARPS_serval_drift}, but for HARPS-DRS derived RVs.}
    \label{fig:HARPS_drs_drift}
\end{figure*}

Figures \ref{fig:HARPS_serval_drift} and \ref{fig:HARPS_drs_drift} show the four RV-$t_{\rm mid}$ correlations: for pre- and post-SERVAL RVs, and pre- and post-DRS RVs. The correlations' significance and slopes are given in the insets. While SERVAL RVs show no correlation in pre RVs, and slightly positive correlation in post RVs, the DRS RVs show a small but significant negative correlation in pre RVs, corresponding to an average nightly drift of $\lesssim-1$\,m\,s$^{-1}$, and no correlation in post RVs.

In many cases the DRS pipeline provides an estimate of the RV drift relative to the nightly calibration sequence, which is calculated by using the light of a reference calibration source injected simultaneously with the star light through a second fibre. As part of deriving the RVs with SERVAL, we corrected the estimated Doppler shifts whenever the drift value was given. The same was done with DRS RVs. Nevertheless, the RV-$t_{\rm mid}$ correlations in pre-DRS and post-SERVAL RVs remain.


We do not know the reason for these correlations, which can be further investigated on a deeper instrumental level. For instance, one can look for correlation with other auxiliary information. For the purpose of this data-driven work, we simply corrected the RVs for the small correlations in all four RV sets. Hence, the final correction model for each RV included both the NZP, according to the night of observation, and the average intra-night drift, according to the $t_{\rm mid}$ of observation. The model uncertainties were added in quadrature to the internal RV uncertainties.

\subsection{\label{sec:RVjump}The 2015 instrumental RV jump}

By the end of May 2015 the HARPS spectrograph underwent a major fibre link upgrade, during which the old circular fibres were replaced by octagonal ones
as described in \citet{LoCurto2015}.
They find that the main improvement was an increase of the scrambling power by at least a factor of ten, and that due to this any de-centring of the stellar light source on the fibre affect the measured RV by less than $0.5$\,m\,s$^{-1}$. 
While this in general is expected to result in more stable RVs, the intervention resulted in a significant change of the instrumental profile. As reported, this introduced an RV offset between the pre- and post-upgrade epochs.
\citet{LoCurto2015} investigated this "jump" by comparing the pre- to post-upgrade RVs for more than 20 RV standard stars, and found RV offsets between $-2.3$\,m\,s$^{-1}$ and $20.0$\,m\,s$^{-1}$. Their data also indicates that the offsets 
might be related to the spectral type of the targets. 
The reason for the jump is likely the missing optimization of the current DRS extraction pipeline. Maybe it can be better calibrated by a future DRS version, but in particular asymmetry changes in the line spread function are challenging to model.
A dependence on spectral type is expected, since the systematics might vary across the detector, and the RV information also changes with spectral type across the detector.

We also investigated the magnitude of the jump with SERVAL. For this purpose, we recomputed again the HARPS RVs, but this time with a common stellar template, meaning that pre- and post-data were processed jointly and not separated. Then we recomputed the NZPs for this entire data set as described in \autoref{subsec:NZPmethod}. 

\begin{figure*}
    \centering
    \includegraphics[width=18cm]{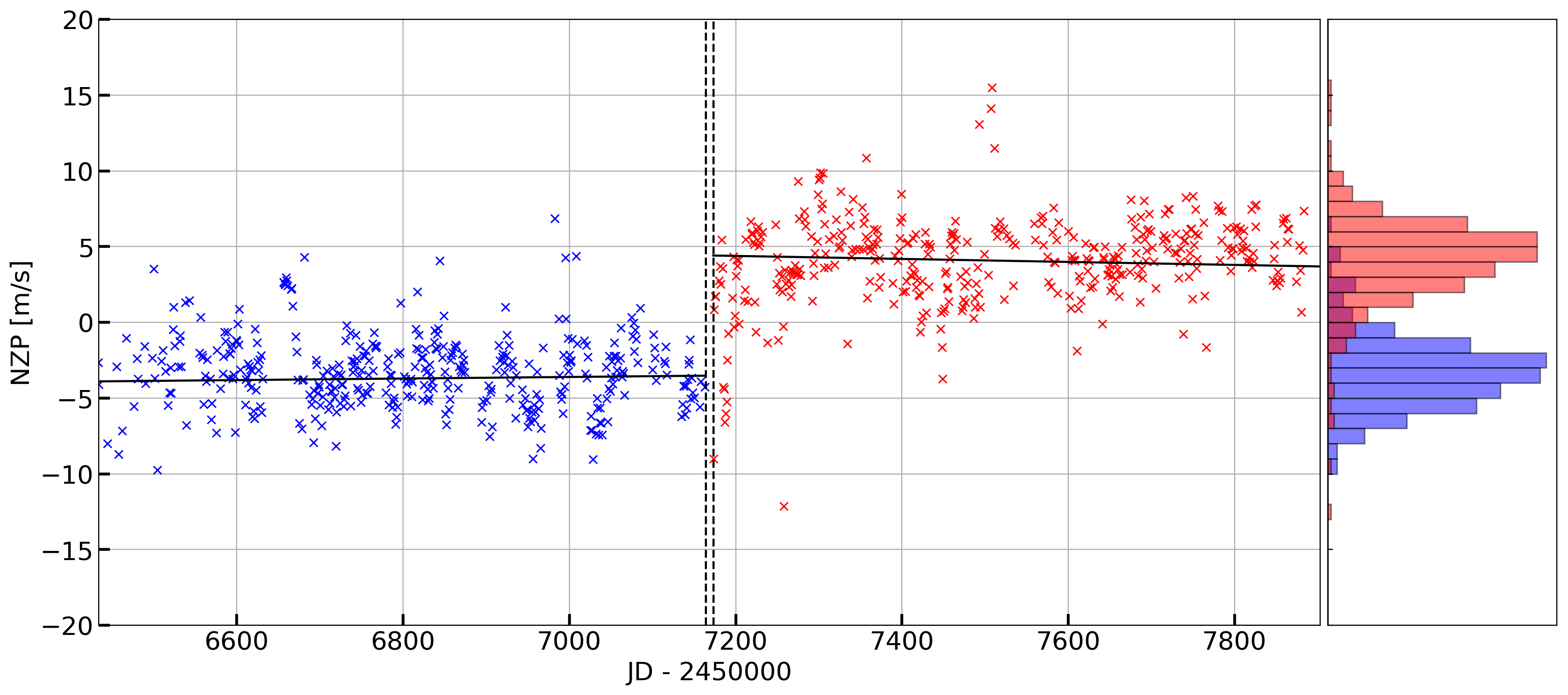}
    \caption{Estimated NZPs from the stripped data set of the SERVAL RVs, zoomed to four years around the fibre upgrade (vertical lines) together with the histogram of their distribution on the side.
    On top of the NZP data points the corresponding linear fits are plotted.}
    \label{fig:HARPS_NZP_jump}
\end{figure*}

\begin{figure}
    \centering
    \includegraphics[width=\linewidth]{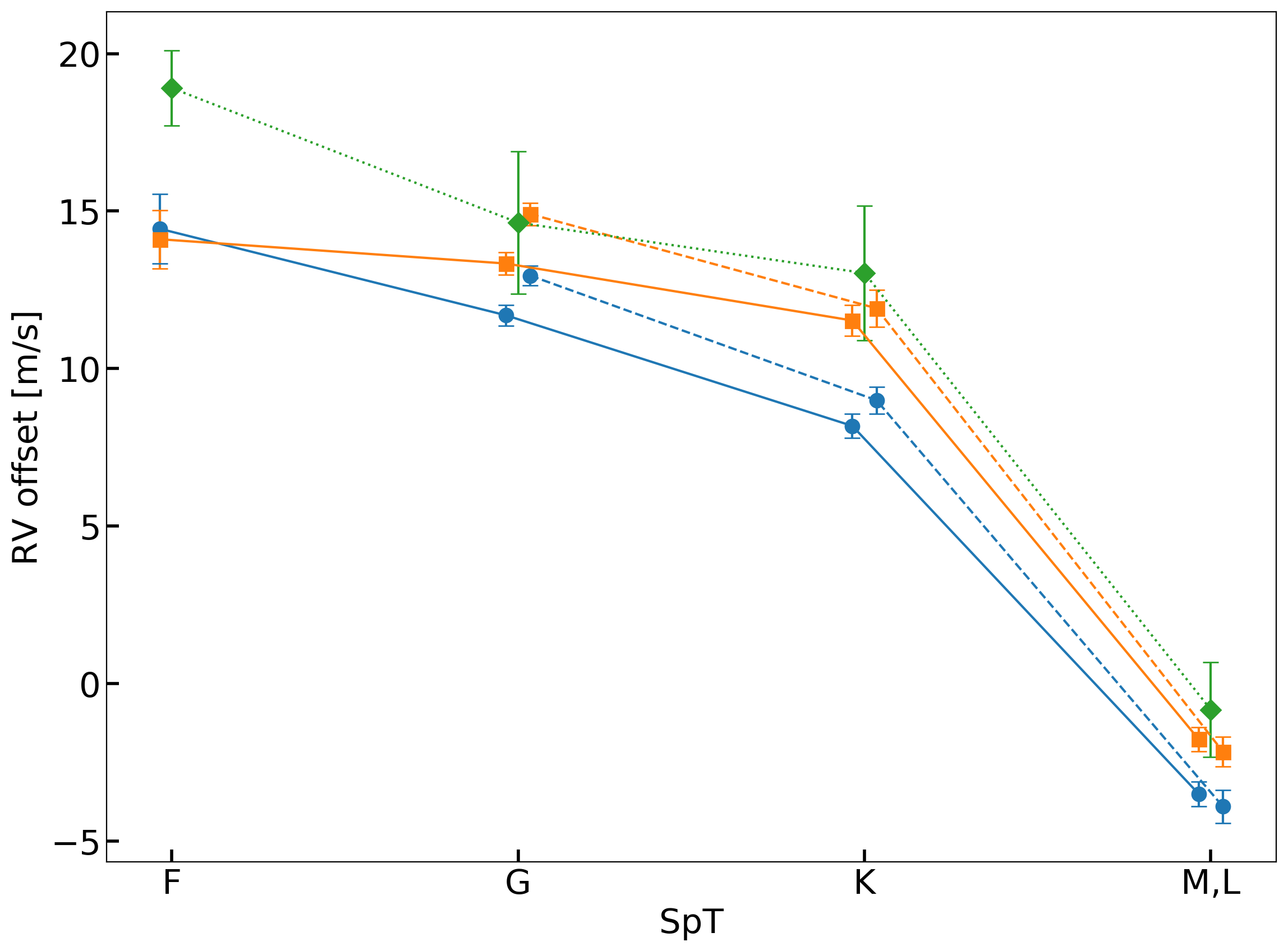}
    \caption{RV jump estimates from all data sets for different spectral types. The two RV pipelines SERVAL (blue circles) and DRS (orange squares) are color-coded. Offsets reported by \citet{LoCurto2015} were grouped by spectral type and averaged for comparison (green diamonds). The dashed lines indicate stripped data sets, while the solid lines correspond to the full sets.}
    \label{fig:jump_sptype}
\end{figure}

Following this procedure, we found that while the computed NZPs were stable for the pre-upgrade epoch, they showed a steady drift for the post-upgrade period. 
Therefore, in order to estimate the RV offset, we applied separate linear fits to the NZPs before and after the intervention break. Only valid NZPs that were computed from nights with sufficient RV data were used for the fits and consequently for the determination of the offset. This procedure, using the entire data set, yielded a jump of $8.19\pm 0.38$\,m\,s$^{-1}$ between the last pre-upgrade night ($\mathrm{JD} = 2\,457\,163$) and the first post-upgrade night ($\mathrm{JD} = 2\,457\,173$)\footnote{A few stars were observed with HARPS during this maintenance time, and consequently their DRS RVs appear in our database. However, we opted not to use these data for the precision analysis and avoid correcting them for NZPs.}.

\begin{table}
    \caption{RV jumps between the pre- and the post-upgrade epochs determined from RVs computed by both pipelines (SERVAL and DRS). Reported are results from the entire data sets, as well as for all sub-samples, depending on spectral type.}
    \label{table:jump_results}
    \centering
    \renewcommand{\arraystretch}{1.2} 
    \begin{tabular}{@{}l c c c c@{}}
        \topline 
        Method    & \multicolumn{2}{c}{full} & \multicolumn{2}{c}{stripped}\\  
        Sp. Type  & $N_\mathrm{quiet}$ & RV offset [m\,s$^{-1}$]& $N_\mathrm{quiet}$ &  RV offset [m\,s$^{-1}$]\\
        \midline 
        \multicolumn{5}{c}{SERVAL}\\ 
        all   &  1516  &  $ 8.19 \pm 0.38$ & 441 &  $ 7.93 \pm 0.33$ \\
        F     &   154  &  $14.44 \pm 1.10$ &  - &  - \\
        G     &   617  &  $11.69 \pm 0.33$ & 192 &  $12.95 \pm 0.31$ \\
        K     &   421  &  $ 8.18 \pm 0.38$ & 133 &  $8.99  \pm 0.43$ \\
        M, L  &   248  &  $-3.51 \pm 0.39$ &  43 &  $-3.91 \pm 0.53$ \\
        \midline 
        \multicolumn{5}{c}{DRS}\\ 
        all   &  1503  &  $10.21 \pm 0.43$ & 379 &  $ 9.35 \pm 0.38$ \\
        F     &   146  &  $14.11 \pm 0.93$ &  - &  - \\
        G     &   611  &  $13.33 \pm 0.36$ & 156 &  $14.90 \pm 0.35$ \\
        K     &   384  &  $11.53 \pm 0.49$ & 102 &  $11.91 \pm 0.59$ \\
        M, L  &   275  &  $-1.77 \pm 0.39$ &  45 &  $-2.17 \pm 0.47$ \\
        \bottomline
    \end{tabular}
\end{table}

\begin{figure}
    \centering
    \includegraphics[width=1.05\linewidth]{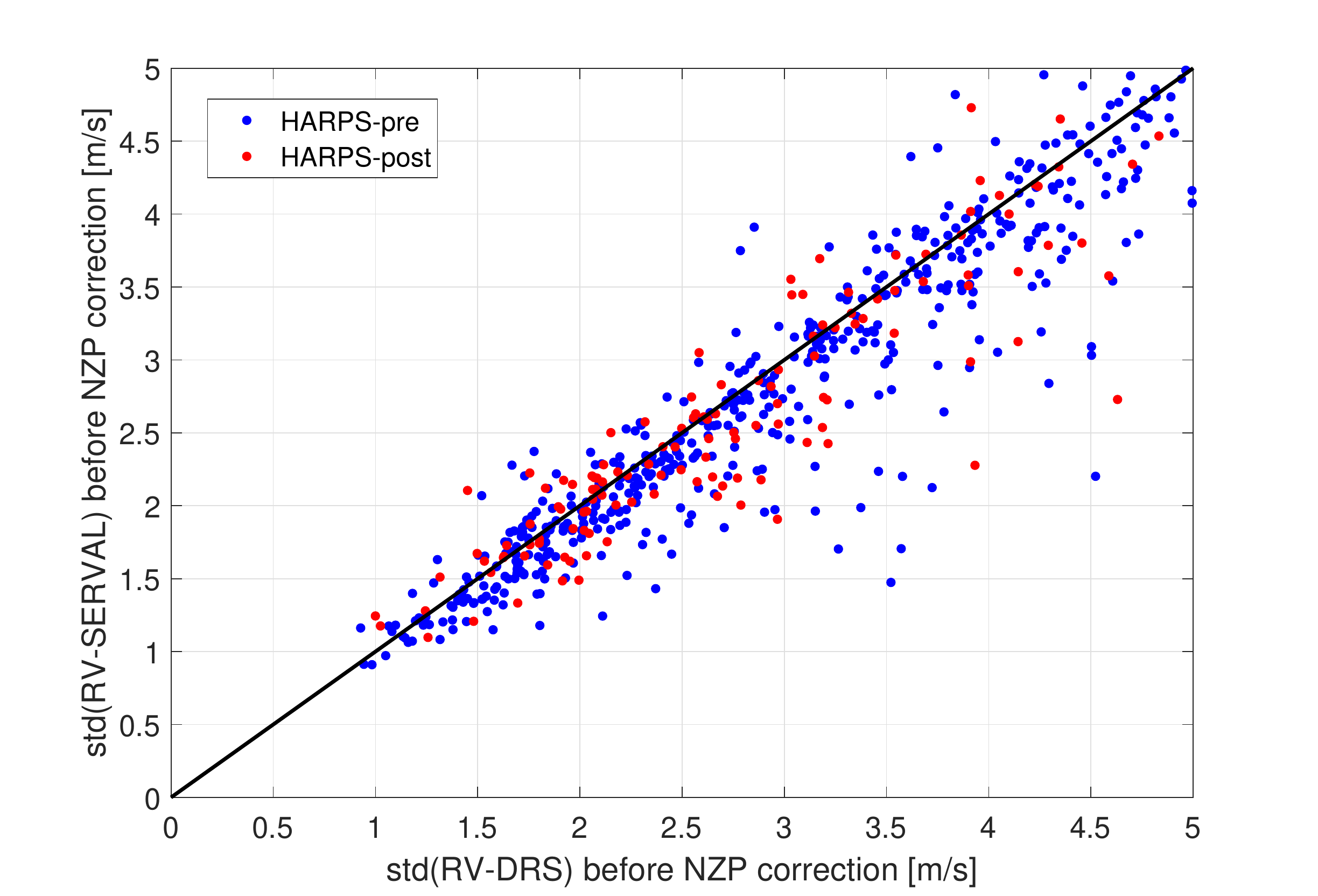}
    \caption{Comparison of SERVAL and DRS-CCF as RV derivation tools. To estimate the std of the RVs of each star, we used wrms. Only stars with $>10$ RVs and an observing time span of $>7$\,d are displayed ($492$ stars from HARPS-pre, and $128$ stars from HARPS-post).}
    \label{fig:drs_serval_std_comp}
\end{figure}

\begin{figure*}
    \centering
    \includegraphics[width=0.45\linewidth]{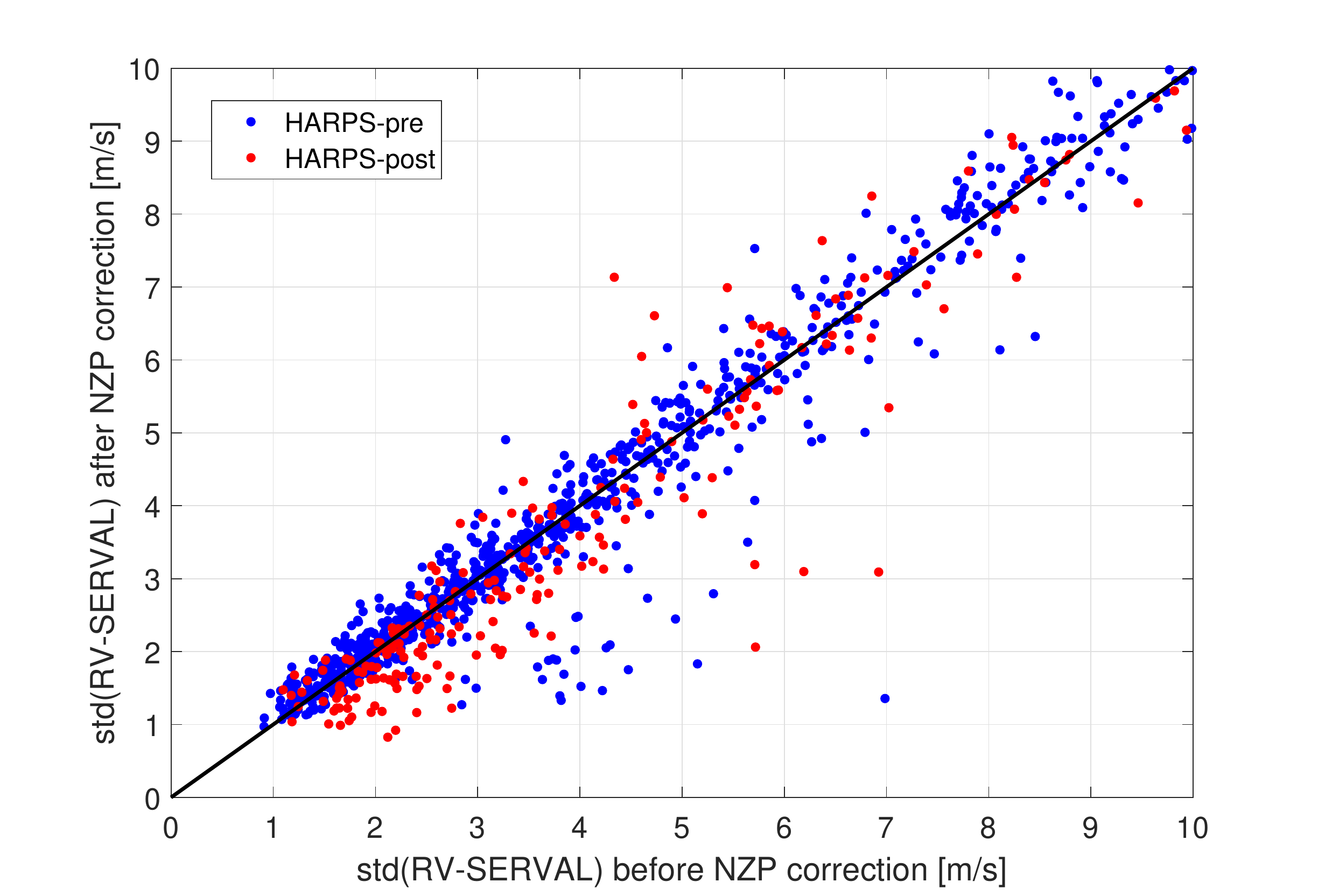}\includegraphics[width=0.45\linewidth]{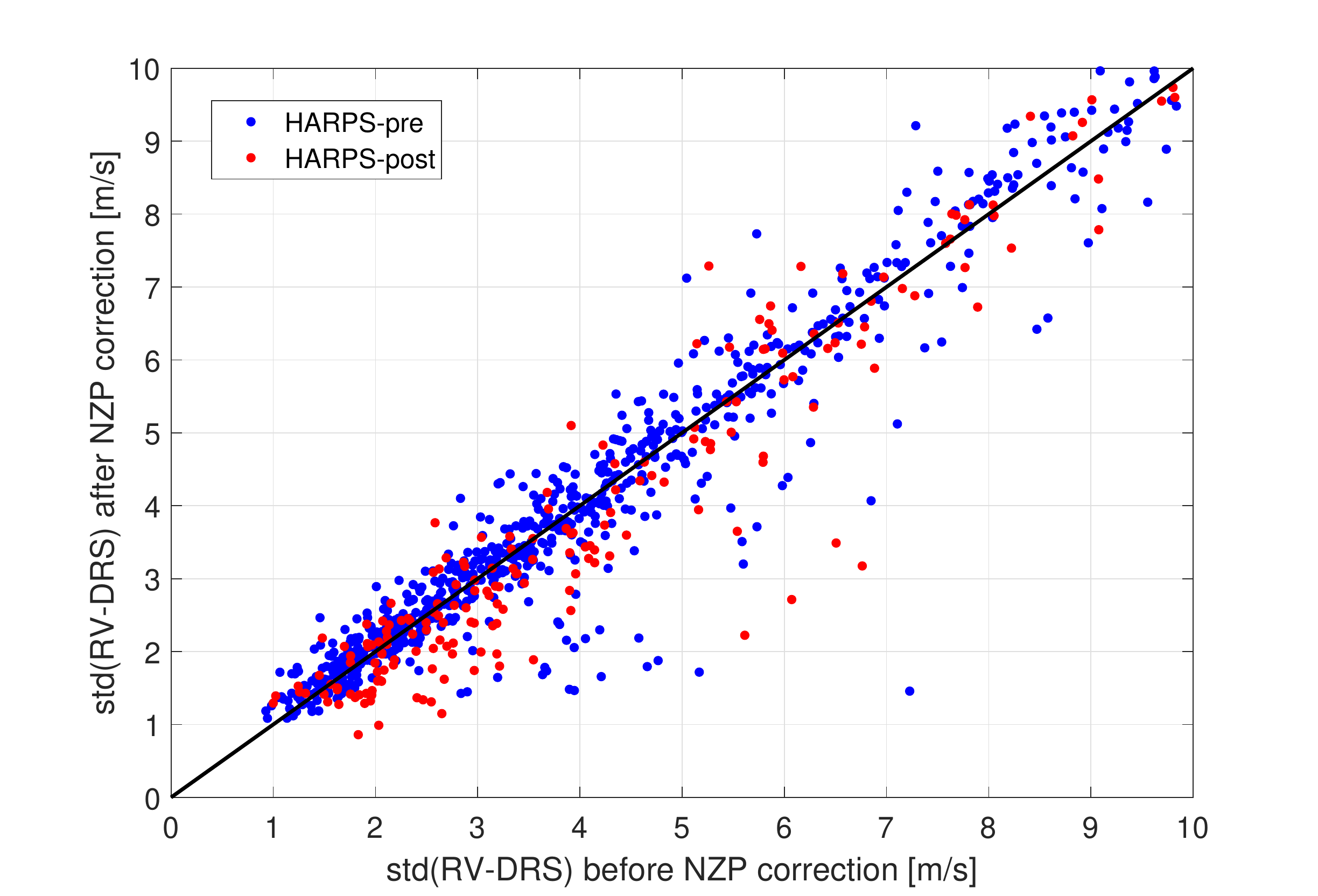}
    \caption{Comparison of the std(RV) per star before and after correcting the RVs for NZP variations. Only stars with $>10$ RVs and an observing time span of $>7$\,d are displayed. \emph{Left}: SERVAL RVs  ($834$ stars from HARPS-pre, and $213$ stars from HARPS-post). \emph{Right}: DRS RVs ($711$ stars from HARPS-pre, and $193$ stars from HARPS-post).}
    \label{fig:HARPS_serval_std_comp}
\end{figure*}

In order to minimize systematics in the offset estimation that might be introduced by any drifts of the NZPs, in particular in combination with unequal distributions in the numbers of observations between the two epochs, we repeated the same procedure (NZP computation and jump determination) with a modified (reduced) data set. This "stripped" set was created by discarding for each star some of the earliest or latest RVs, so that the numbers of nights in which the stars were observed, were equal for the two epochs. The NZPs estimated from this stripped data are illustrated in \autoref{fig:HARPS_NZP_jump}. The plot is clipped to a time span of four years around the instrumental intervention.
The jump in the NZPs between nights prior to the upgrade and those after is evident and is also reflected in the bimodality of the NZP distribution. 
For the stripped data we found an RV offset of $7.93\pm 0.33$\,m\,s$^{-1}$, which is consistent with the first approach. The reported errors are propagated from the linear fits, where the uncertainties of the regressions were estimated by bootstrapping, for which the data was re-sampled 1000 times.
For comparison we repeated the entire analysis also with the DRS RVs. With offsets of
$10.21\pm 0.43$\,m\,s$^{-1}$ for the entire data set, and $9.35\pm 0.38$\,m\,s$^{-1}$ for the stripped data set, they appear to show higher jumps than those determined from the SERVAL RVs. 

We also investigated the dependence of the magnitude of the RV jump on spectral type. To do so, we gropued the stars by their spectral type, and repeated the analysis on each sub-sample alone. The results from all analyses are summarized in \autoref{table:jump_results}.

For the sub-samples of later spectral types the offset estimates from both methods (full data and stripped data) are similar and consistent within their uncertainties, while they deviate a bit for spectral type G. No jump values could be estimated for the stripped data sets of the F-type stars.
This is due to the combination of a small number of stars and the reduced data set, which together lead to insufficient number of nights for which the NZPs could be estimated. However, they were determined for the full data sets. The dependence on the spectral type is evident. The offsets are highest for early type stars and appear to decrease for lower effective temperatures. For M~dwarfs the sign of the jump is inverted. The apparent relationship is illustrated in \autoref{fig:jump_sptype}, where all offset estimates from the SERVAL and DRS RVs are plotted. The shift between the SERVAL and the DRS results is visible, and is highest for K stars. Still, the trend with spectral type is consistent across all different data sets. Also, our jump estimates from the DRS data are consistent with those determined by \citet{LoCurto2015}.

The resulting offsets should be treated with care, as it is expected that the jumps vary from one source to another. However, the results presented here are still a valuable source of information. They can either be used for sanity checks, when offsets between pre- and post-upgrade RV data are treated as free parameters in RV modeling, as usually done in planet searches, or they can even serve as priors for the offset. In particular for large amplitude signals, such priors can serve as useful constraints.

\section{Results}
\label{Sec5}

\subsection{The new HARPS RV database}
\label{HARPS_database}

\citet{Butler2017} published an example of a well documented, user-friendly, high-precision RV database. Their database contains a large collection of precise Doppler measurements, derived from $\sim 65\,000$ spectra of $\sim1700$ stars, as well as stellar-line activity-index measurements, obtained with the iodine cell method \citep{Marcy1992, Valenti1995, Butler1996} between 1996 and 2014 with the KECK-HIRES\footnote{Similarly to the ESO archive, publicly available spectra obtained with the HIRES can be found in the Keck Observatory Archive at \url{https://koa.ipac.caltech.edu/cgi-bin/KOA/nph-KOAlogin},
but to our knowledge these data do not contain final-product RV measurements.} spectrograph \citep{Vogt1994}. The RVs published by \citet{Butler2017} were by far the most precise and extensive HIRES-RV archive available to the exoplanet community. These public HIRES RVs were the basis for a number of new exoplanet discoveries and orbital updates \citep[e.g.][]{Butler2017, Trifonov2018A&A...612A..49R, Kaminski2018, Trifonov2019b, Tuomi2019}, and form an important RV validation archive for the {\it TESS} candidates in the  Northern hemisphere. The large sample size of the HIRES RV archive has also allowed us to identify and correct the data for systematic variations, by calculating nightly zero point variations, and an average intra-night drift \citep{Tal-Or2019MNRAS.484L...8T}.

\begin{figure*}
    \centering
    \includegraphics[width=0.45\linewidth]{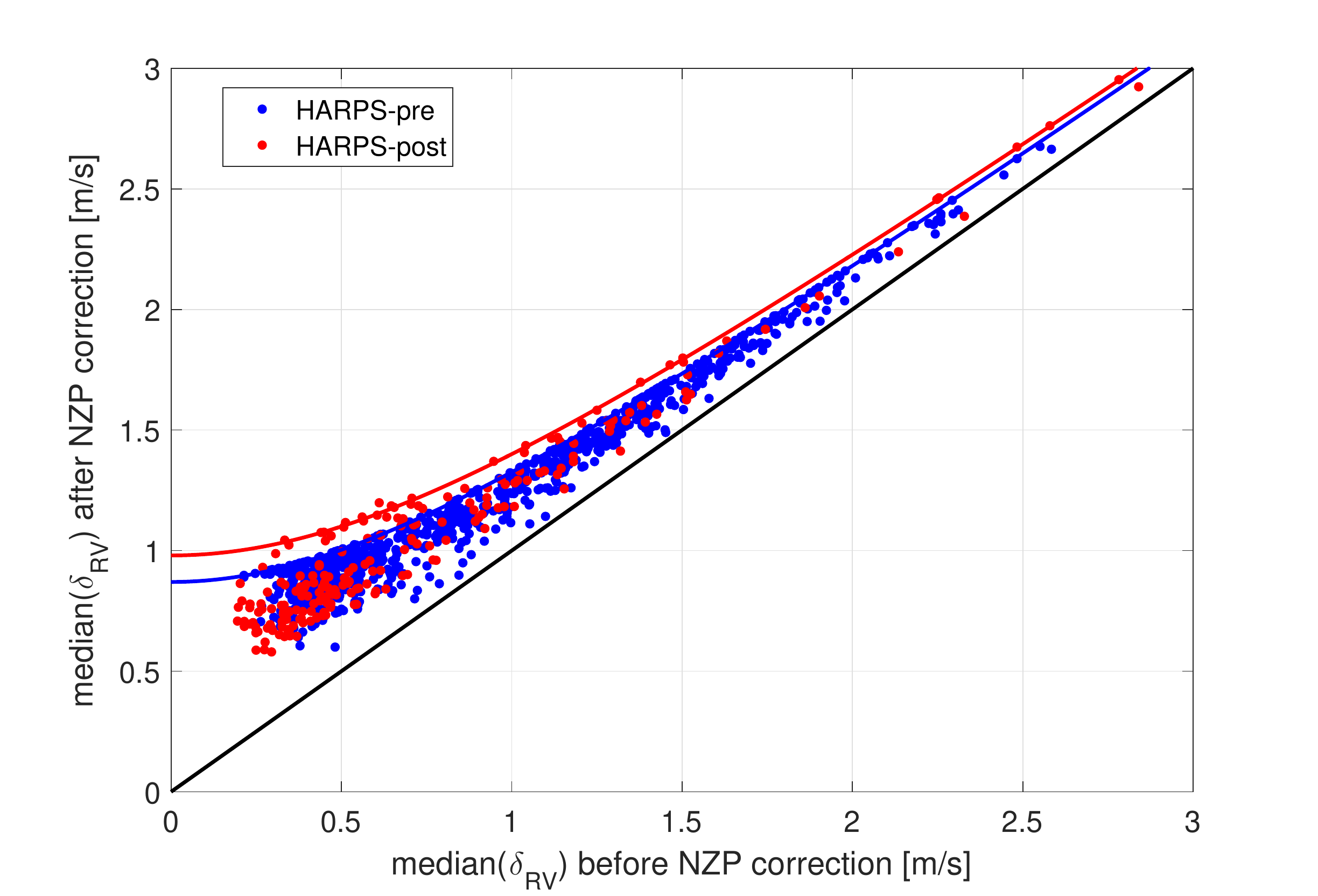}\includegraphics[width=0.45\linewidth]{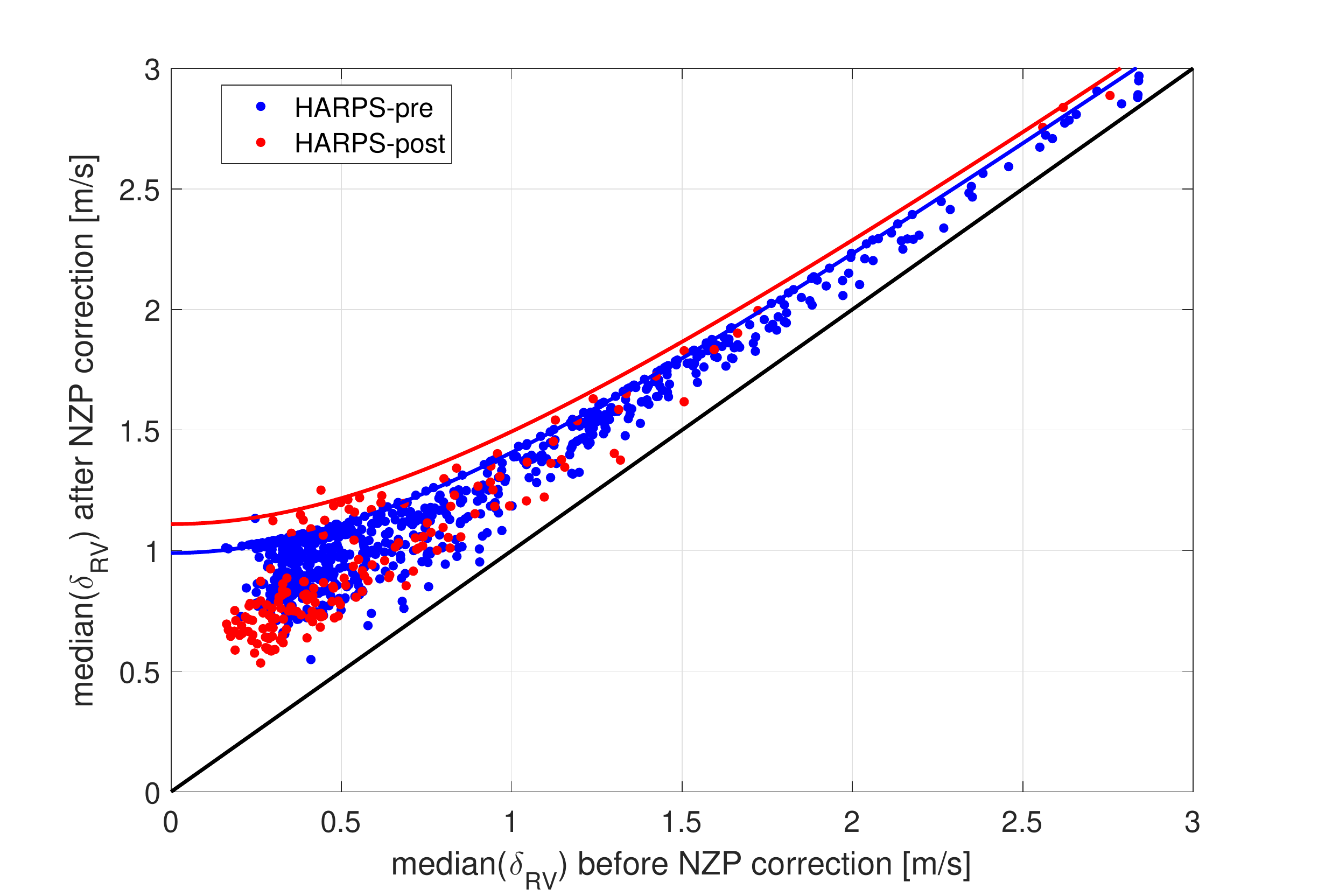}
    \caption{The impact of NZP error propagation on the median RV uncertainty per star (med($\delta{\rm R}$)). \emph{Left}: SERVAL RVs. \emph{Right}: DRS RVs. The solid lines demonstrate the impact of adding max($\delta$NZP) as listed in \autoref{table:nzp_stats}.}
    \label{fig:HARPS_serval_erv_comp}
\end{figure*}

Following a similar route to \citet{Butler2017}, we created {\sc HARPS-RVB}ank -- a public database based on the results presented in this work. {\sc HARPS-RVBank} is available on CDS\footref{note1} or on its official web page\footnote{\url{www.mpia.de/homes/trifonov/HARPS_RVBank.html}} and provides up-to date SERVAL and DRS data products for the HARPS targets. 
\autoref{tab:RVs} shows an excerpt of the database with some important columns. The {\sc HARPS-RVB}ank  provides original as well NZP-corrected SERVAL and DRS RV measurements, their BJD epoch, activity index measurements, and uncertainties.
From the DRS products, the user can find the CCFs FWHM, contrast, and the BIS-span measurements. From the SERVAL spectral analysis, we provide the chromatic index (CRX), the differential line width (dLW), and the H$\alpha$, Na I D, and Na II D activity-related line measurements.
We note that the activity time series are also affected by the 2015 fibre upgrade (see \autoref{sec:RVjump}) leading to a notable post-upgrade 
"jump" in the data.
We recommend to treat the pre- and the post-upgrade
activity time series as taken from two different instruments, 
before testing the data for the presence of periodic signals.

The new archive also provides all applied individual RV corrections to the data, such as barycentric Earth radial velocity (BERV), secular acceleration of the RV \citep[SA;][]{Kurster2003}, Fabry-Perot (FP) drift, DRS and SERVAL NZP time series, and the final correction value of each RV, including the average intranight drift.
Additionally, we provide for each epoch auxiliary observational information such as 
exposure time, S/N of the spectra at 550\,nm, quality flag, type of DRS binary mask used, 
principal investigator (PI) and ESO program-ID of the observation.

For user flexibility we also provide a {\sc github} repository\footnote{\url{https://github.com/3fon3fonov/HARPS_RVBank}} 
of the {\sc HARPS-RVBank} database, where the user can find the final products, 
useful tools, and instructions. There we also provide a MySQL, and a JavaScript Object Notation (JSON) versions of the {\sc HARPS-RVBank} database, which could be easily integrated as an object storage in modern programming languages such as Python, or used in web interfaces.

\subsection{\label{subsec:NZP_impact}DRS versus SERVAL and the impact of the NZP correction on RV-quiet stars}

Figure~\ref{fig:drs_serval_std_comp} compares SERVAL and DRS-CCF as RV derivation tools. For the comparison, we focus on the most quiet stars (RV scatter $<5$\,m\,s$^{-1}$), and compute the average (and median) ratio std(RV-SERVAL)/std(RV-DRS), where we use wrms as our std estimator. We find that for HARPS-pre SERVAL yields RVs with an average wrms improvement of $\sim5$\% ($\sim4$\% median), while for HARPS-post the average wrms improvement is $\sim4$\% ($\sim2$\% median). Hence, for most stars SERVAL yields a slightly better RV precision, and should in general be preferred over the nominal DRS RVs.

Figure~\ref{fig:HARPS_serval_std_comp} shows the impact of the NZP correction on the wrms of all RV-quiet stars (RV scatter $<10$\,m\,s$^{-1}$). Comparing the impact of the correction between HARPS-pre and HARPS-post, we see that the average impact on HARPS-pre data is smaller than the average impact on HARPS-post data, for both SERVAL and DRS RVs. In order to quantify the effect, we again look at the average (and median) wrms improvement due to the NZP correction. For both SERVAL and DRS RVs, the NZP correction yields only a marginal effect on HARPS-pre data ($<2$\%), while for HARPS-post data the average reduction of wrms is $\sim8$\% ($\sim6$\% median). This finding is consistent with the fact that, for both RV-derivation pipelines, HARPS-post NZPs have larger scatter and smaller uncertainties than HARPS-pre NZPs (see \autoref{table:nzp_stats}), leading to a more significant correction of post RVs with the calculated NZPs. 


\begin{figure*}
    \centering
    \includegraphics[width=0.45\linewidth]{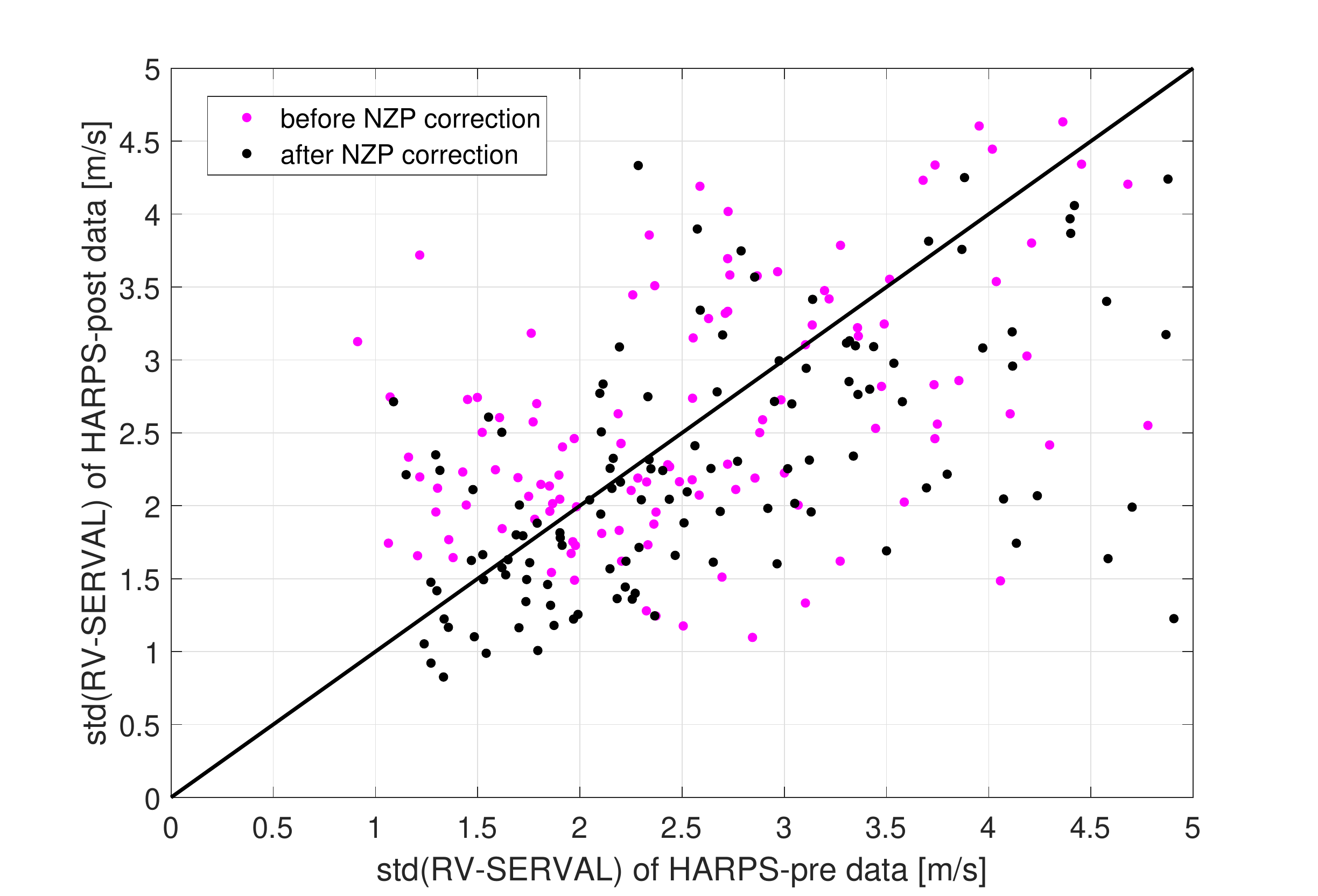}\includegraphics[width=0.45\linewidth]{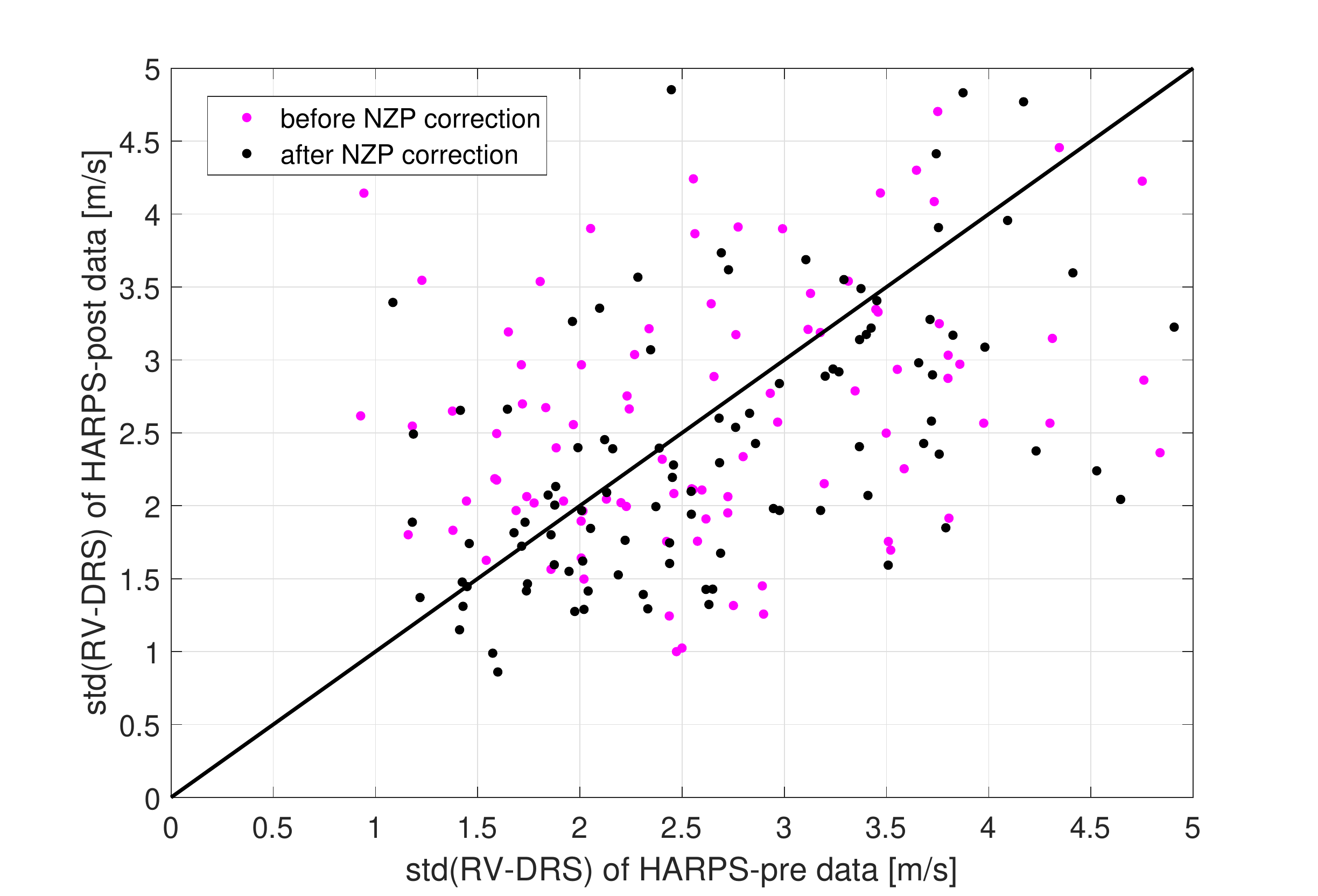}
    \caption{Comparison of std(RV) per star before and after the $2015$ intervention. Only stars with $>10$ RVs and an observing time span of $>7$\,d are displayed. \emph{Left}: SERVAL RVs ($115$ stars before NZP correction, and $120$ stars after it). \emph{Right}: DRS RVs ($93$ stars before NZP correction, and $99$ stars after it).}
    \label{fig:pre-post_serval_std_comp}
\end{figure*}

\begin{figure*}
    \centering
    \includegraphics[width=0.45\linewidth]{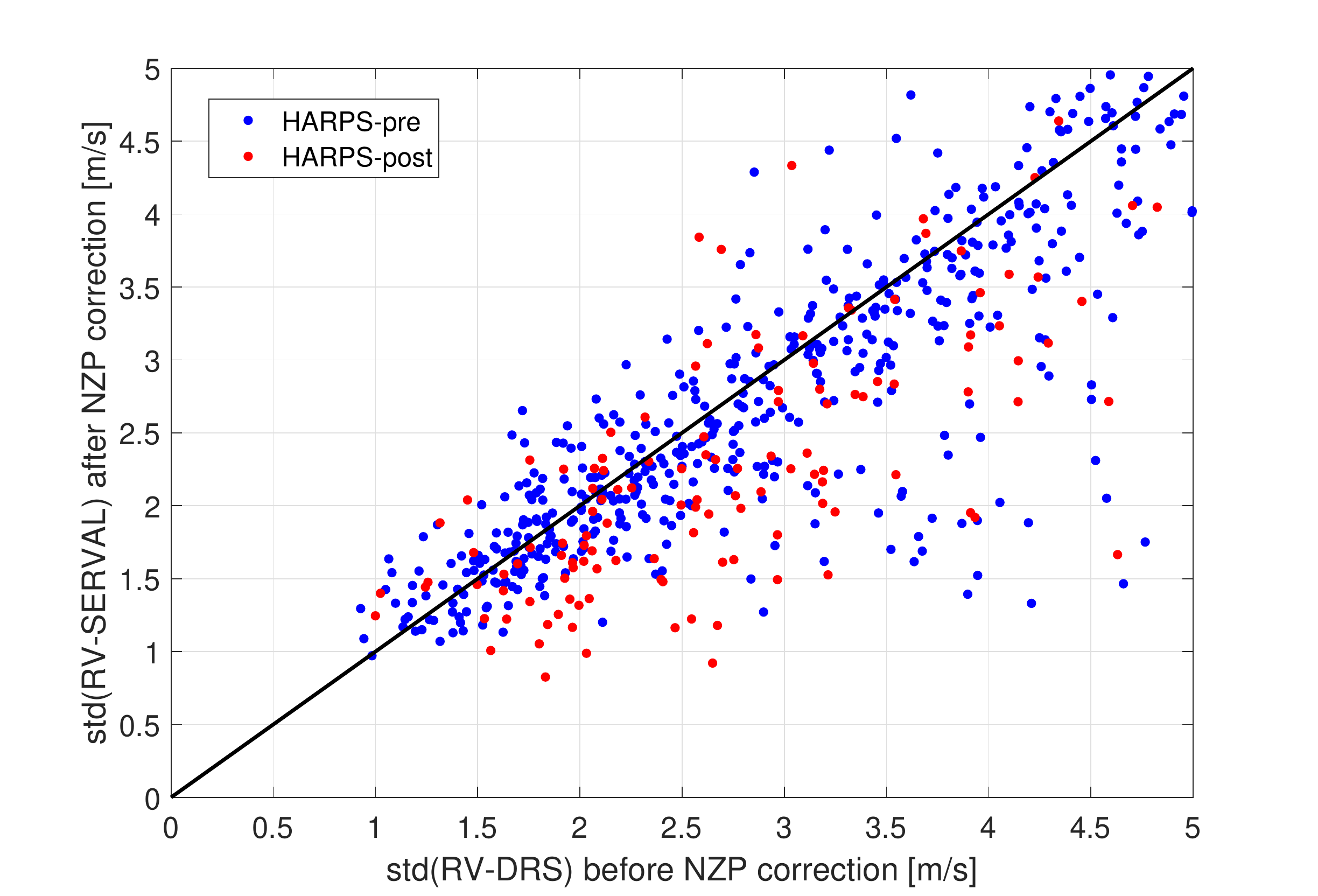}\includegraphics[width=0.45\linewidth]{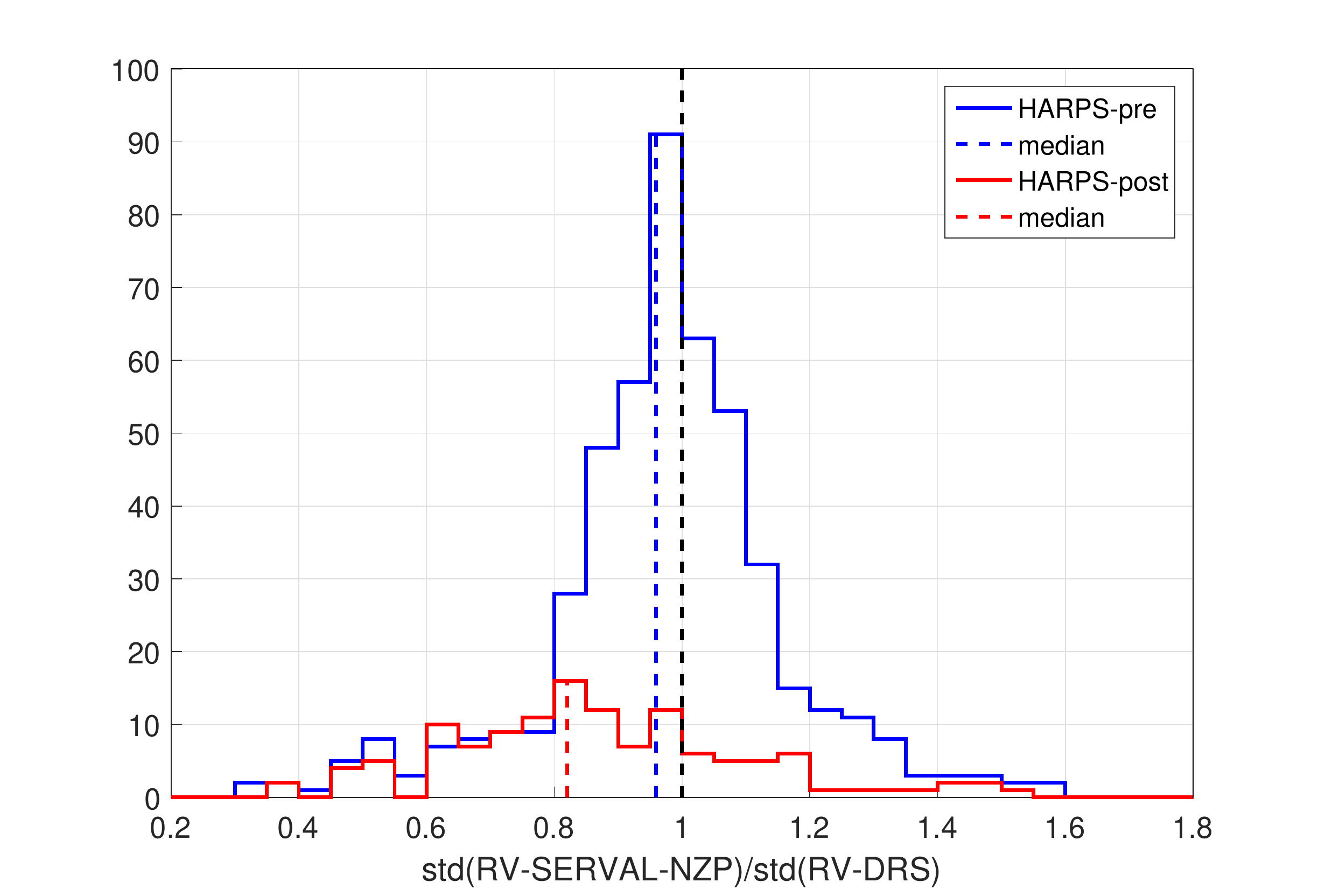}
    \caption{The combined impact of extracting HARPS RVs with SERVAL and correcting for the NZPs. Only stars with $>10$ RVs and an observing time span of $>7$\,d are displayed ($485$ stars from HARPS-pre, and $126$ stars from HARPS-post).}
    \label{fig:drs_serval-nzp_std_comp}
\end{figure*}

Interestingly, in HARPS-pre there is a larger number of stars with a correction that significantly deviates from the mean. Specifically, in HARPS-pre there are $\sim20$ stars with a wrms improvement of $\gtrsim2$\,m\,s$^{-1}$, while there are only four such stars in HARPS-post. This effect is probably related to the larger number of significantly outlying NZPs in HARPS-pre (see \autoref{table:nzp_stats}). For instance, there are two stars in HARPS-pre with a wrms reduction from $>5$\,m\,s$^{-1}$ to $<2$\,m\,s$^{-1}$: HD\,145927 and HD\,197818. HD\,145927 has $10/37$ of its RVs corrected by outlier NZPs of $\sim-5$\,m\,s$^{-1}$, and HD\,197818 has one RV corrected by an outlier NZP of $\sim-30$\,m\,s$^{-1}$, which occurred on the night of JD\,$=2\,456\,246$. 
In HARPS-post, three of the four stars with a wrms improvement of $\gtrsim2$\,m\,s$^{-1}$ (HD\,16417, HD\,20781, and HD\,197536) have at least one observation on the night of JD\,$=2\,457\,258$, 
which has an outlier NZP of $\sim-14$\,m\,s$^{-1}$. The fourth star (GJ\,506) has many exposures on the night of JD\,$=2\,457\,174$, right after the fibre link upgrade.

Figure~\ref{fig:HARPS_serval_erv_comp} shows the impact of the NZP correction on the median RV uncertainty per star (med($\delta{\rm RV}$)). Co-adding the NZP uncertainties ($\delta$NZPs) to the original RV uncertainties enlarged the med($\delta{\rm RV}$) values. However, the estimated $\delta$NZPs rarely exceeded $\sim1$\,m\,s$^{-1}$. We believe that the new RV uncertainties better represent HARPS' RV precision, and will require a smaller jitter term in modeling the RV time series with Keplerian orbits. Moreover, the nights with the largest number of bright RV-quiet stars, observed under good conditions, naturally give the smallest $\delta$NZPs. This, in turn, gives higher weight to the RVs from the nights in which we have a good estimate of the calibration errors.


Another demonstration of the importance of the NZP correction can be seen in \autoref{fig:pre-post_serval_std_comp}, which compares the RV wrms per star before and after the $2015$ intervention.  
Without NZP correction, HARPS' fibre exchange actually worsened the wrms per star by $\sim10$\% on average. However, after correcting for the NZPs, we find it to improve the wrms per star by $\sim10$\% on average. Moreover, the pre and post wrms values are more correlated when correcting for the NZPs. 
This indicates that the fibre exchange indeed improved the instrument performance, but 
the wrms values of the most quiet stars were still dominated by instrumental systematic effects.
It is the NZP correction that enables the better precision achieved with HARPS' new fibres.

Focusing again on the most quiet stars (RV scatter $<5$\,m\,s$^{-1}$), \autoref{fig:drs_serval-nzp_std_comp} shows end-to-end comparison between the RV performance of the nominal DRS-CCF (without NZP correction) and SERVAL with NZP correction. For HARPS-pre data, the average wrms improvement is $\sim5$\% ($\sim4$\% median), which is dominated by a small number of stars with a relatively large wrms improvement. For HARPS-post data, the average wrms improvement is $\sim15$\% ($\sim18$\% median), with only a few stars above the 1:1 line. The difference can be explained by the different NZP behaviors of HARPS-pre and HARPS-post RVs. We conclude that for HARPS spectra the NZP-corrected SERVAL RVs are in general more precise than the DRS RVs, and we regard them as the main product of this work.

\subsection{A practical example: Orbital update of the GJ\,253 multi-planet system}

We now make a direct comparison of the official HARPS-DRS RV data, and our final product --- the SERVAL NZP-corrected RV data, by testing their overall quality when modeling an actual planetary system discovered with HARPS. For this purpose we use the known multi-planet system GJ\,253 (HD\,51608), whose RVs are consistent with two Neptune-mass planets \citep{Mayor2011, Udry2019}. We have selected this system rather arbitrarily, as in our HARPS archive we find many known planetary systems, which could serve as better (or worse) examples for testing our SERVAL-NZP data. The GJ\,253 system, however, is 
appropriate in the context of the results presented in this work, because:

\begin{itemize}
    \item The DRS data have no outliers (e.g. wrong user RV-guess), meaning that we can perform a full data set comparison with the SERVAL data.
    \item Data were taken both before (218 RVs) and after (9 RVs) the May 2015 fibre upgrade. 
    \item GJ\,253 has a sufficient number of archival RV data for a proper statistical comparison. 
    \item The planetary signals are strong and unambiguous (i.e. likely real, and not due to stellar activity). 
    \item The system is fairly complex (a two-planet system), but is still easy to analyze, since there is no need of N-body modeling.
    \item We can perform an update of the systems' orbital solution. 
    
\end{itemize}

\begin{figure}[tp]
    \centering
    \includegraphics[width=9cm]{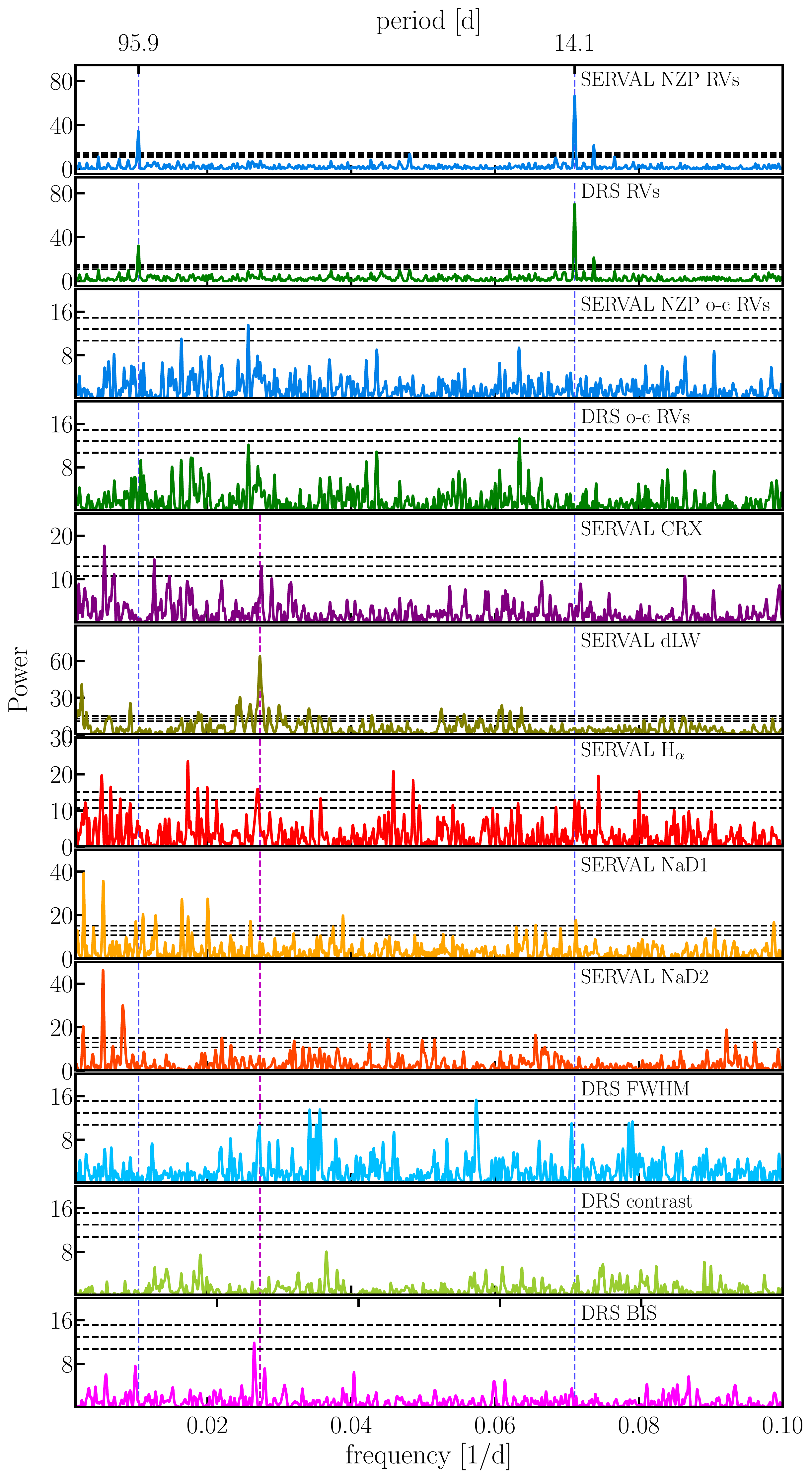} 
 
    \caption{GLS power spectrum for the GJ\,253 data, based on NZP-corrected SERVAL RVs, DRS RVs, and stellar activity indicators from DRS and SERVAL as labelled in the panels. The horizontal lines in the GLS periodograms show the FAP levels of 10\%, 1\%, and 0.1\%. Blue vertical lines indicate the orbital period of GJ\,253 b \& c.
    The (magenta vertical line indicates the periodicity near 36\,d, which is present in DRS FWHM, BIS-span, contrast, and the SERVAL dLW activity indicators likely related to the stellar rotational period. 
    }
    \label{RV_results} 
\end{figure}

\autoref{RV_results} shows GLS periodogram analyses of the DRS and SERVAL RVs, and activity index time series of the HARPS spectra of GJ\,253.
Both data sets are clearly consistent with significant power at periods of 14.07\,d and 96.11\,d, which are the 
reported planetary periods of GJ\,253 b and c.
There is no evidence of significant GLS power in the activity indices from DRS and SERVAL, that could be 
associated with the two RV signals. The only exception is the SERVAL's NaD\,I activity indicator, 
which seems to show a marginally significant power near 14.04\,d.
This NaD\,I peak, however, is only the tenth strongest peak in the
NaD\,I GLS power spectrum. 
A closer inspection of the RV and the NaD\,I time series, showed that these two peaks are sufficiently distant ($>$3$\sigma$) in
frequency space, and show no correlation.
Therefore the strong RV signal, and the marginally 
significant NaD\,I signal are likely not related to each other.\looseness=-8

For the fitting analysis of the GJ\,253 system we use {\em The Exo-Striker} fitting toolbox\footnote{{\em The Exo-Striker} is freely available at \url{https://github.com/3fon3fonov/exostriker}} \citep{Trifonov2019_es}.
To identify the planetary signals embedded in the data, we perform a "blind-search" using the {\em The Exo-Striker's} "Auto fit" algorithm, which, as its name suggests, automatically inspects the RV data for periodic signals via GLS periodogram search and performs pre-whitening signal subtraction \citep{Hatzes2013}. Finally, when no significant peaks are left in the RV data residual periodogram {\em The Exo-Striker} performs a subsequent simultaneous best-fit optimization of the planetary
semi-amplitude $K$, the orbital period $P$, eccentricity $e$, argument of periastron $\omega$, and mean anomaly $M$ at the first observational epoch.
We consider the HARPS-pre and HARPS-post fibre upgrade data as obtained from separate instruments, and thus we optimize their RV offsets and RV velocity jitter \citep{Baluev2009} simultaneously with the planetary parameters. Using this approach we were able to instantly identify the planetary signals published in the discovery paper by \citet{Mayor2011}, in both the DRS data and the SERVAL NZP-corrected data.
As a next step, {\em The Exo-Striker} employs a Markov Chain Monte Carlo \citep[MCMC, via the $emcee$ sampler;][]{emcee}  sampling around the best fit, and we adopt the 1$\sigma$ confidence level of the posterior distributions as parameters uncertainties.

\begin{figure*}[tp]
    \centering
    \includegraphics[width=9cm]{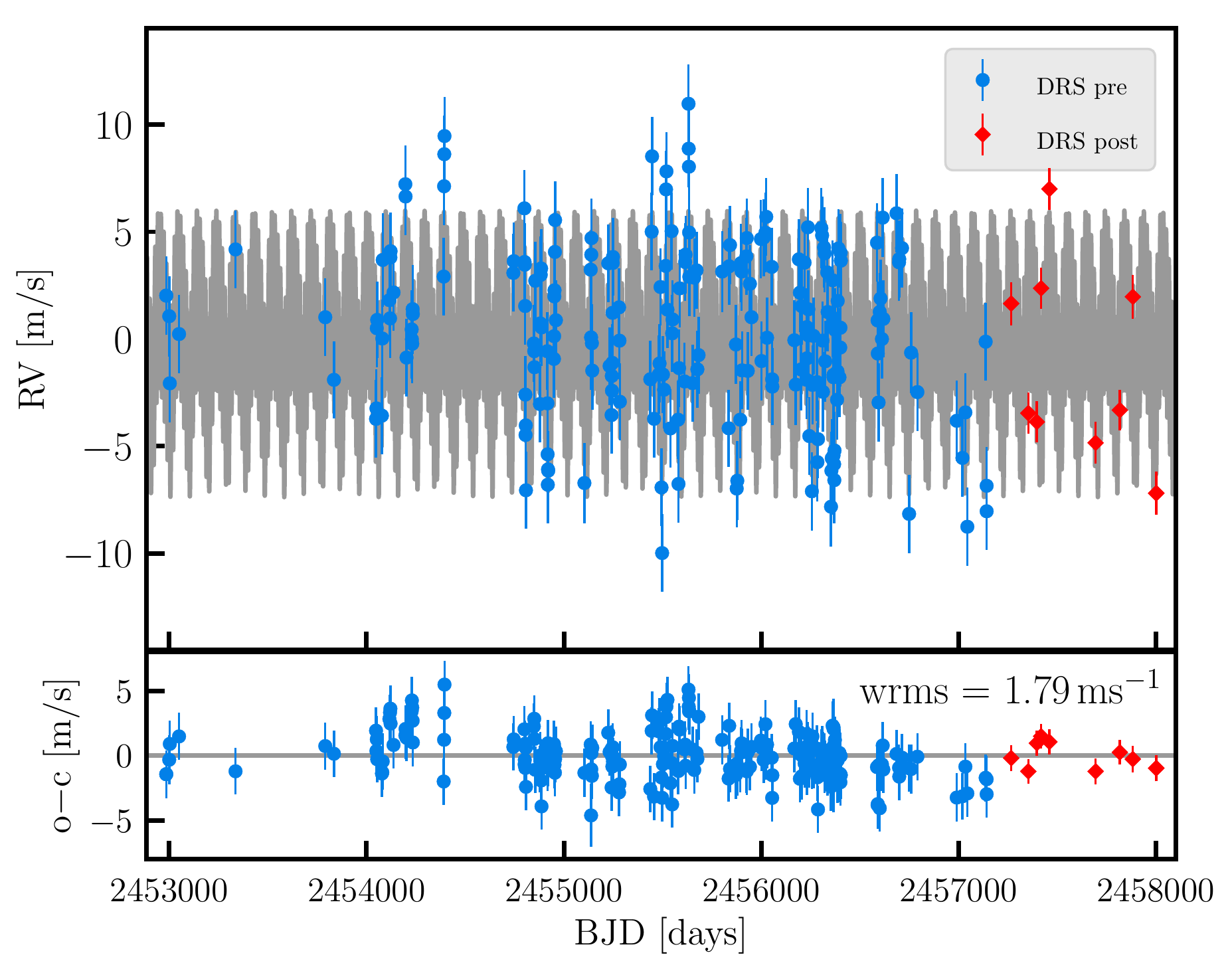}
    \includegraphics[width=9cm]{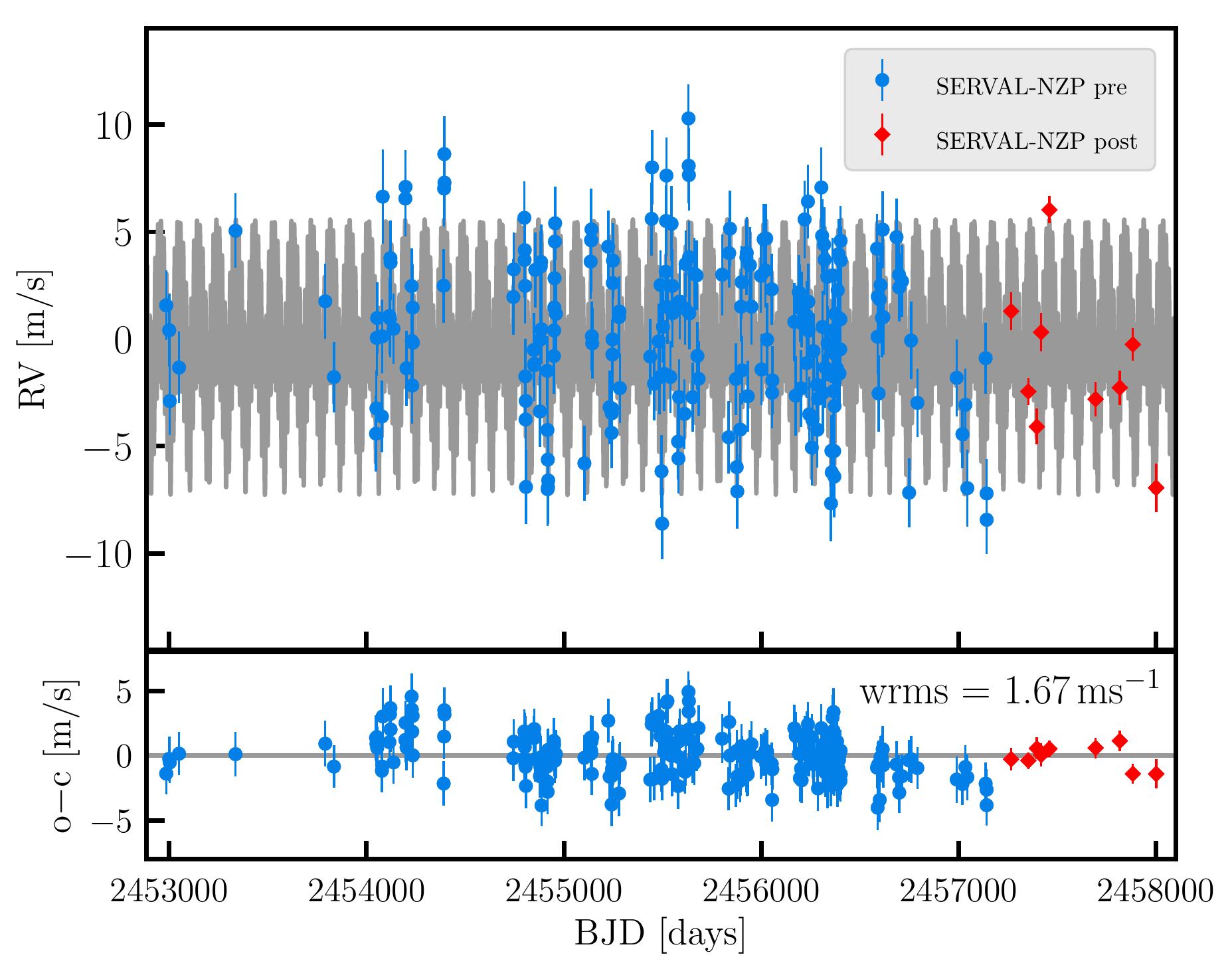}
    \includegraphics[width=8.5cm]{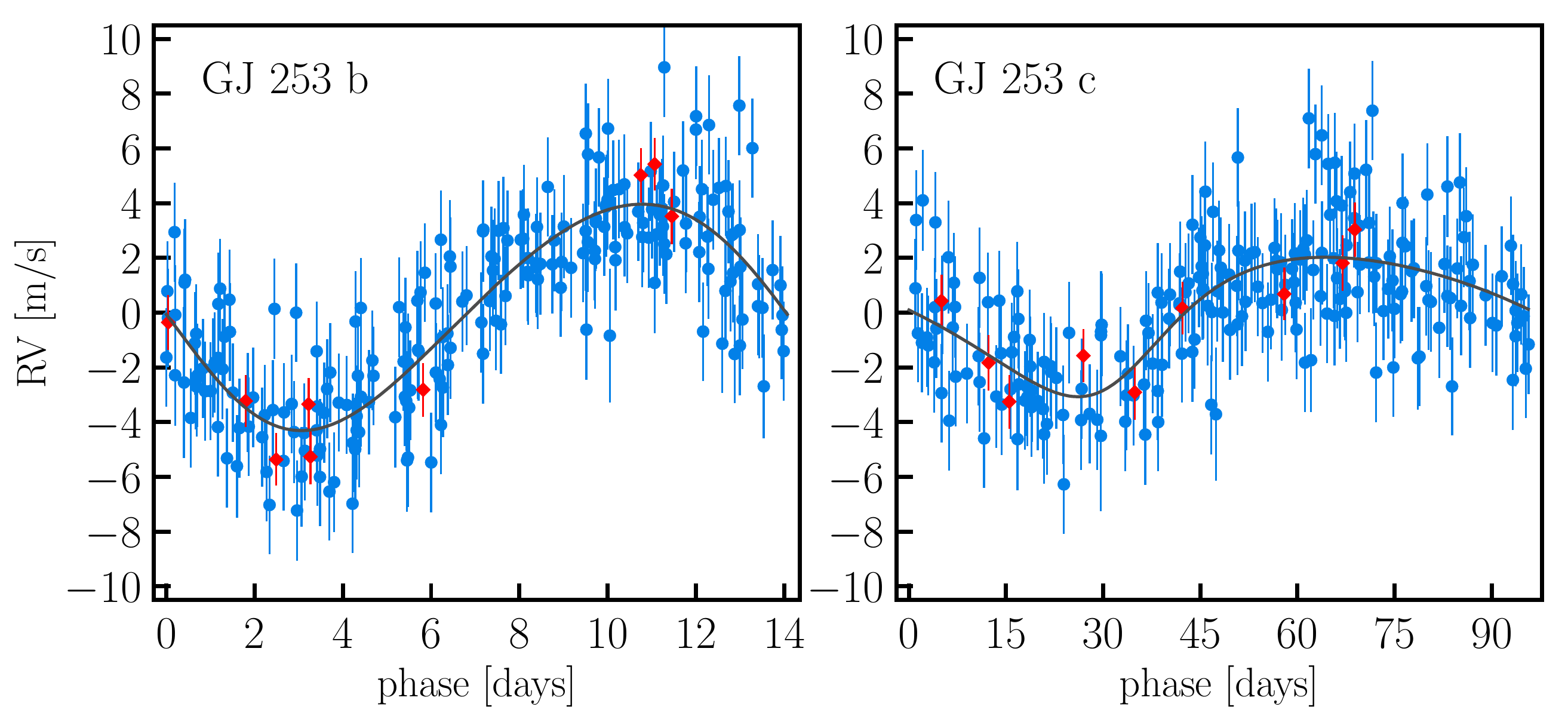}
    \hspace{0.5cm}
    \includegraphics[width=8.5cm]{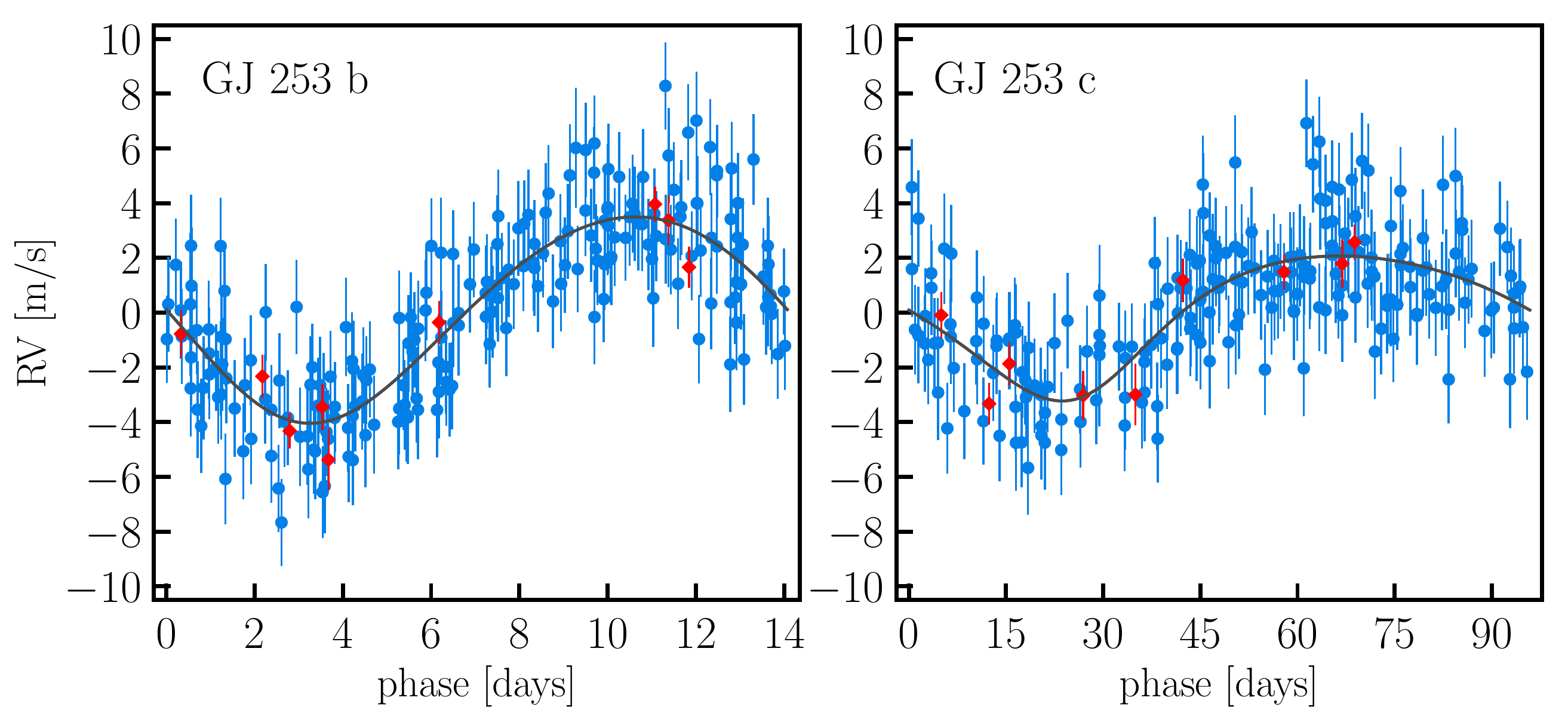}\\
    \caption{
HARPS RVs of GJ\,253 derived with DRS (\emph{left}) and SERVAL (\emph{right panel}) and modelled with a two-planet model. Blue data (218 RVs) are taken before the May 2015 fibre upgrade (pre), while red data (9 RVs) are after (post). \emph{Lower panels}: RVs phase folded to the Doppler signals of GJ\,253 b and c.}
    \label{RV_results2} 
\end{figure*}

\autoref{table:best-fits} summarizes the best-fit parameters and MCMC derived asymmetric 
uncertainties from our dual RV analysis, while 
\autoref{RV_results2} shows the best two-planet fits of the DRS and SERVAL NZP-corrected RV data sets.
Both data sets lead to consistent best-fit planetary periods of $P_{\rm b} \sim$ 14.07 days and $P_{\rm c} \sim$ 95.9\,d, which are in agreement with the period estimates by \citet{Mayor2011}, who gives 
$P_{\rm b}$ = 14.070$\pm$0.004 days and  $P_{\rm c}$ = 95.42$\pm$0.39 days.
We derive lower and better constrained eccentricities when compared to those provided in the discovery work of \citet{Mayor2011}.
The DRS data leads to 
$e_{\rm b}$ = 0.087$_{-0.056}^{+0.030}$  and $e_{\rm c} $ =  0.258$_{-0.099}^{+0.059}$, SERVAL NZP-corrected data suggest
$e_{\rm b}$ = 0.091$_{-0.061}^{+0.027}$ and $e_{\rm c}$ = 0.227$_{-0.074}^{+0.050}$, 
while \citet{Mayor2011} gives $e_{\rm b}$ = 
0.15$\pm$0.06 and $e_{\rm c}$ = 0.41$\pm$0.18.
We note that \citet{Mayor2011} had used only a little more than half of the HARPS data we use in our analysis, and thus it is not surprising that our estimates are better constrained by the larger, and longer baseline data set.\looseness=-2

\citet{Udry2019} have recently published an orbital update of the GJ\,253 system.
They based their orbital solution and activity analyses only on HARPS spectra obtained before the fibre upgrade,
but with data reduction done using a new version of the HARPS-DRS pipeline.
The MCMC-based best-fit solution given in \citet{Udry2019}
is generally consistent with our best-fit estimates (see Table\,10 in their paper).


To derive the planetary minimum masses and semi-major axes, we adopt the recent stellar mass estimates for GJ\,253
by \citet{Soto2018}, who used HARPS spectra to infer stellar parameters and derived a stellar mass of $M$ = 0.85 $M_\odot$. 
The DRS data and the SERVAL-NZP corrected data mutually agree on the RV semi-amplitudes of GJ\,253 b and c, 
and thereafter on the derived minimum masses of $m_{\rm b} \sin i \sim 0.04M_{\rm jup}$ and $m_{\rm c} \sin i \sim 0.05M_{\rm jup}$. 
Thus, our orbital update analysis is consistent with two warm Neptune-mass planets in orbit around GJ\,253, as it was reported by \citet{Mayor2009}, and \citet{Udry2019}.

    \begin{table*}[ht]
    
    \centering   
    \caption{{Orbital parameters of the two-planet system GJ\,253 derived from DRS and SERVAL-NZP data.}}   
    \label{table:best-fits}      
    
    \begin{tabular}{lrrrrrrrr}     
    
    \hline\hline  \noalign{\vskip 0.7mm}      
    
\makebox[0.1\textwidth][l]{\hspace{62 mm}DRS data fit      \hspace{51.5 mm} SERVAL-NZP data fit \hspace{1.5 mm} } \\
\cline{2-3}\cline{5-6}\noalign{\vskip 0.9mm}   
    
    Parameter \hspace{25.0 mm} & GJ\,253 b & GJ\,253  c  & \hspace{25.0 mm} & GJ\,253 b & GJ\,253 c \\
    \hline \noalign{\vskip 0.7mm} 
    
        $P$  [d]                      &     14.073$_{-0.001}^{+0.002}$ &     95.889$_{-0.088}^{+0.116}$  &&  14.071$_{-0.001}^{+0.002}$ &     95.870$_{-0.055}^{+0.112}$  \\ \noalign{\vskip 0.9mm}
        $K$  [m\,s$^{-1}$]            &     4.14$_{-0.23}^{+0.12}$ &      2.55$_{-0.25}^{+0.17}$   &&  3.77$_{-0.16}^{+0.17}$ &      2.65$_{-0.26}^{+0.10}$ \\ \noalign{\vskip 0.9mm}  
        $e$                           &     0.087$_{-0.056}^{+0.030}$ &   0.258$_{-0.099}^{+0.059}$   &&  0.081$_{-0.056}^{+0.027}$ &      0.232$_{-0.072}^{+0.052}$\\ \noalign{\vskip 0.9mm}
        $\omega$  [deg]               &    118.6$_{-30.6}^{+48.9}$ &    217.5$_{-21.8}^{+15.5}$   &&  151.7$_{-42.4}^{+45.4}$ &    200.9$_{-19.0}^{+18.2}$ \\ \noalign{\vskip 0.9mm}
        $M_{\rm 0}$  [deg]            &   234.8$_{-47.5}^{+30.7}$ &    254.6$_{-16.4}^{+25.6}$  &&  190.8$_{-43.2}^{+46.2}$ &    269.6$_{-15.9}^{+23.0}$ \\ \noalign{\vskip 0.9mm}

        $a$  [au]                     &      0.10806$_{-0.00001}^{+0.00001}$  &      0.38838$_{-0.00024}^{+0.00031}$   &&  0.10805$_{-0.00001}^{+0.00001}$  &      0.38833$_{-0.00015}^{+0.00032}$ \\ \noalign{\vskip 0.9mm} 
        $m \sin i$  [$M_{\rm jup}$]   &      0.0439$_{-0.0026}^{+0.0012}$   &      0.0498$_{-0.0046}^{+0.0031}$   &&  0.0400$_{-0.0017}^{+0.0018}$  &      0.0521$_{-0.0049}^{+0.0019}$  \\ \noalign{\vskip 1.5mm} 
        
        $\gamma_{\rm HARPS-pre}$~[m\,s$^{-1}$]             & \makebox[\dimexpr(\width-5em)][l]{39977.20$_{-0.12}^{+0.15}$} & &  \makebox[\dimexpr(\width-12em)][l]{-0.23$_{-0.11}^{+0.14}$}                &  \\ \noalign{\vskip 0.9mm}   
        $\gamma_{\rm HARPS-post}$ [m\,s$^{-1}$]             &  \makebox[\dimexpr(\width-5em)][l]{39991.06$_{-0.51}^{+0.40}$}  & & \makebox[\dimexpr(\width-12em)][l]{0.41$_{-0.46}^{+0.30}$} & \\ \noalign{\vskip 0.9mm}
        
        $\sigma_{\rm HARPS-pre}$ [m\,s$^{-1}$]              &      \makebox[\dimexpr(\width-5em)][l]{1.77$_{-0.04}^{+0.16}$} &&  \makebox[\dimexpr(\width-12em)][l]{1.46$_{-0.04}^{+0.17}$} &\\ \noalign{\vskip 0.9mm}
        $\sigma_{\rm HARPS-post}$ [m\,s$^{-1}$]             &     \makebox[\dimexpr(\width-5em)][l]{0.90$_{-0.03}^{+0.86}$} &&  \makebox[\dimexpr(\width-12em)][l]{0.37$_{-0.01}^{+0.92}$} &\\ \noalign{\vskip 0.9mm}
    \hline \noalign{\vskip 0.7mm} 
        wrms [m\,s$^{-1}$]        &      \makebox[\dimexpr(\width-5em)][l]{1.79}  & & 
        \makebox[\dimexpr(\width-12em)][l]{{1.67}} & \\        
        $-\ln\mathcal{L}$             &    \makebox[\dimexpr(\width-5em)][l]{454.00} &&  \makebox[\dimexpr(\width-12em)][l]{{437.544}} &\\
        $\Delta\ln\mathcal{L}$             &    \makebox[\dimexpr(\width-5em)][l]{180.00} &&  \makebox[\dimexpr(\width-12em)][l]{{185.99}} &\\

    \hline \noalign{\vskip 0.7mm}

    \end{tabular}  
    
    
    \tablefoot{\small All orbital elements are Jacobi orbital elements, and are valid for the first HARPS observational epoch, which is BJD = 2\,452\,984.733}
    
    \end{table*}

The orbital dynamics of the GJ\,253 system is beyond the scope of this work, but for the sake of completeness we also examine the long-term orbital stability of the 
best-fit configurations achieved form the different data sets. 
For this purpose, we adopt the Wisdom-Holman \citep[also known as MVS;][]{Wisdom1991} 
N-body integrator implemented in the {\em The Exo-Striker}, which includes General  Relativistic (GR) precession correction term. We find that given the small orbital separations and significant planetary eccentricities observed in the GJ\,253 system, the GR precession effects on the orbital dynamics are significant, and thus the GR correction  must be included in the long-term evolution of the system to assure realistic dynamical outcome.
Overall, a crude stability analysis shows that the GJ\,253 system is stable for at least 10 M\,yr, exhibiting an interesting apsidal alignment libration $\Delta\varpi$ = $\varpi_{\rm b}$ - $\varpi_{\rm c}$ $\sim$ 0$^\circ$ with a semi-amplitude of 60$^\circ$, and significant oscillations of the planetary eccentricities in the range 0.02 $ < e_{\rm b} < $ 0.17 and 0.21 $ < e_{\rm c} < $ 0.23 with mean values of $\overline{e_{\rm b}}$ = 0.11, and $\overline{e_{\rm c}}$ = 0.22, respectively. The time scale of the secular orbital oscilation is $\approx$ 37\,000 yr.

In \autoref{RV_results} it is evident that the SERVAL dLW, CRX, 
and $H_\alpha$ activity indicators suggest a periodicity near 36.5\,d. It is particularly strong 
in the SERVAL dLW, where this periodicity has a significant power. 
A peak near this period is also detected in the DRS BIS-span, and the DRS FWHM time series.
A peak at a similar frequency 
was also detected by \citet{Udry2019} on their HARPS activity time series.
We are likely witnessing stellar spots rotating with the rotational period of the star.  
Interestingly, the RV residuals in both DRS and SERVAL show a strong, but insignificant GLS peak near 38.9\,d. 
While this peak seems sufficiently distant from the $\sim$ 36.5-d period seen in the 
activity index time series, it is still possible that it could be related to the stellar rotation.

Finally, when it comes to the quality of the two fits, the DRS-CCF data yields $\Delta\ln\mathcal{L}$ = 180.00, 
with respect to the null hypothesis (i.e. no planets), while the SERVAL-NZP has $\Delta\ln\mathcal{L}$ = 185.99, which means that the SERVAL-NZP corrected RV data adds significant evidence in favour of the two planets. This also manifests in a lower weighted rms of the SERVAL-NZP residuals compared to DRS-CCF (see \autoref{table:best-fits}).
Therefore, our orbital update on the GJ\,253 system is based on the two-planet Keplerian modeling of the SERVAL NZP-corrected data.

This practical example of the GJ\,253 system shows that our SERVAL data are indeed a better choice
with respect to the official DRS data. 
It is still possible that for other systems the DRS data would lead to better fits, but overall, given the sample statistics comparison given in \autoref{subsec:NZP_impact} (see also Figs. \ref{fig:drs_serval_std_comp} and \ref{fig:drs_serval-nzp_std_comp}),
we are confident that the NZP-corrected SERVAL data should be preferred in most cases. 
In particular for stars of late spectral type such as K, M and L, we expect SERVAL to outperform the DRS-CCF.

\section{Summary and conclusions}
\label{Sec6}

In this paper we present an independent systematic analysis of the HARPS spectral archive.  In particular, we re-calculated Doppler velocity measurements of the publicly accessible spectra, and  performed stellar line analysis, which is important for validating planetary induced Doppler signals. 
For this part of the analysis, we applied the SERVAL RV pipeline, with which we derived slightly more 
precise RV measurements when compared to those derived by the official ESO-DRS pipeline. 
We find that for a sub-sample of stars with a very small RV scatter 
($<5$\,m\,s$^{-1}$), SERVAL RVs are more precise than DRS RVs by $\sim5\%$, on average.

We make all of our HARPS results publicly available, as a service to the exoplanet community. We provide original uncorrected DRS and SERVAL pipeline data, and self-corrected (for NZPs and average intra-night drifts) DRS and SERVAL RVs. All relevant results of our study are made public 
in the user friendly database "HARPS-RVBank", which is available on the {\sc cds}, {\sc github}, or as a stand alone webpage where the user can browse for data. 

This makes our HARPS-RVBank the first easy-to-access publicly available HARPS RV archive, which to our knowledge contains the most precise RV data products to date. 

 

Another objective of our work was to study whether HARPS Doppler measurements suffer from systematic errors, which could bias the orbital parameter estimates of small planets, or worse, induce spurious planetary discoveries.
We find that despite being, with no doubt, a {\it state-of-the-art} RV instrument, HARPS also suffers from small but significant systematic errors of the instrumental zero-point. 
Our NZP analysis reveals stochastic zero-point variations of $\sim1$\,m\,s$^{-1}$, smooth zero-point variations, with a magnitude of $\sim1$\,m\,s$^{-1}$ and a typical timescale of few weeks, and a few dozen nights whose NZPs significantly deviate from the general zero-point trend by few m\,s$^{-1}$ (and up to $30$\,m\,s$^{-1}$). In addition, we find small ($\lesssim1$\,m\,s$^{-1}$) but significant intra-night drifts in DRS RVs before the $2015$ intervention and in SERVAL RVs after it.
The HARPS NZPs systematic errors are likely related to the non-stability of the daily
wavelength calibration \citep{Dumusque2018}.
The next DRS version will likely have an improved wavelength calibration \citep{Coffinet2019},
which to a large degree might resolve these systematic errors.
Then, it would be interesting to recompute the NZPs as a quality check of the 
new DRS wavelength calibration scheme.

Correcting HARPS RVs for the systematic's model, we find additional improvement of the RV scatter, mainly for observations after the $2015$ intervention. Considering the combined effect of deriving the RVs with SERVAL, and correcting them for the small systematic errors, we find the average wrms improvement of pre RVs to be $\sim5$\%, and of post RVs to be $\sim15$\%. For a small number of stars, whose observations were most affected by the significantly-deviating NZPs, we find a much more significant wrms improvement, by a factor $\gtrsim2$.

Investigating the RVs of a sub-sample of RV-quiet stars that were observed both before and after the $2015$ HARPS optical fibre upgrade, we find a discontinuous jump in their absolute RVs that is independent of the RV derivation software (i.e. we find similar results with DRS and SERVAL). Similarly to \citet{LoCurto2015}, we find the jump to be strongly dependent on the spectral type of the target: from $\sim 14$\,m\,s$^{-1}$ for late F-type stars, to $\sim -3$\,m\,s$^{-1}$ for late M dwarfs. 

 As a demonstration of the new data quality, we provide new orbital estimates of the GJ\,253 multi-planet system based on our new HARPS-SERVAL NZP-corrected data, updating
the planetary minimum masses and orbital elements. 
 Similarily to \citet{Udry2019}, we show that  the GJ\,253 b \& c orbits are probably less eccentric than was previously estimated, when fewer RV data were available.
This shows that it is important to update the orbital elements of known planetary systems
when more data 
are accumulated, as this might remove possible higher-eccentricity biases, for example. 
This is especially valid for multi-planet systems. In those systems the eccentricity is an important parameter that can determine their dynamical properties and might shed some light on their formation and evolution.

The HARPS-RVBank is a valuable data source for planet search, re-analysis of known planetary systems, and validation of newly discovered transiting planets. For its better precision, we strongly recommend using the NZP corrected SERVAL RVs, which is the main product of this work.

\begin{acknowledgements}
 We thank the anonymous referee for thorough reading of the manuscript and thoughtful comments that helped to substantially improve this paper.
We are deeply grateful to the HARPS team at Observatoire de Gen\`{e}ve, 
Observatoire de Haute-Provence, Laboratoire d'Astrophysique de Marseille,
Service d'A\'{e}ronomie du CNRS,
Physikalisches Institut de Universit\"{a}t Bern,
ESO La Silla, and
ESO Garching, who built and maintained the HARPS instrument,
and were generous enough to make the data public. Without their seminal work and consistent effort, this study could not have been performed.
We are also extremely grateful to all the PIs and observers 
associated with the ESO programmes listed below.
We respect their hard work, and we hope that they will benefit from our recomputed data.
T.T. thanks to Thomas Henning and Martin K\"urster for support during this study.
L.T. and S.Z. acknowledge support from the Israel Science Foundation (grant no. 848/16).
M.Z. is supported by the Deutsche Forschungsgemeinschaft under DFG RE 1664/12-1 and Research Unit FOR2544 “Blue Planets around Red Stars”, project no. RE 1664/14-1.
T.T. acknowledges support by Bulgarian National Science Programme "Young Scientists and Postdoctoral Candidates 2019".
This work is based on observations collected at the European Organization for Astronomical Research in the Southern Hemisphere under ESO programmes:
0100.C-0097,
0100.C-0111,
0100.C-0414,
0100.C-0474,
0100.C-0487,
0100.C-0750,
0100.C-0808,
0100.C-0836,
0100.C-0847,
0100.C-0884,
0100.C-0888,
0100.D-0444,
0100.D-0717,
0101.C-0232,
0101.C-0274,
0101.C-0275,
0101.C-0379,
0101.C-0407,
0101.C-0516,
0101.C-0829,
0101.D-0717,
0102.C-0338,
0102.D-0717,
0103.C-0548,
0103.D-0717,
060.A-9036,
060.A-9700,
072.C-0096,
072.C-0388,
072.C-0488,
072.C-0513,
072.C-0636,
072.D-0286,
072.D-0419,
072.D-0707,
073.A-0041,
073.C-0733,
073.C-0784,
073.D-0038,
073.D-0136,
073.D-0527,
073.D-0578,
073.D-0590,
074.C-0012,
074.C-0037,
074.C-0102,
074.C-0364,
074.D-0131,
074.D-0380,
075.C-0140,
075.C-0202,
075.C-0234,
075.C-0332,
075.C-0689,
075.C-0710,
075.D-0194,
075.D-0600,
075.D-0614,
075.D-0760,
075.D-0800,
076.C-0010,
076.C-0073,
076.C-0155,
076.C-0279,
076.C-0429,
076.C-0878,
076.D-0103,
076.D-0130,
076.D-0158,
076.D-0207,
077.C-0012,
077.C-0080,
077.C-0101,
077.C-0295,
077.C-0364,
077.C-0530,
077.D-0085,
077.D-0498,
077.D-0633,
077.D-0720,
078.C-0037,
078.C-0044,
078.C-0133,
078.C-0209,
078.C-0233,
078.C-0403,
078.C-0751,
078.C-0833,
078.D-0067,
078.D-0071,
078.D-0245,
078.D-0299,
078.D-0492,
079.C-0046,
079.C-0127,
079.C-0170,
079.C-0329,
079.C-0463,
079.C-0488,
079.C-0657,
079.C-0681,
079.C-0828,
079.C-0927,
079.D-0009,
079.D-0075,
079.D-0118,
079.D-0160,
079.D-0462,
079.D-0466,
080.C-0032,
080.C-0071,
080.C-0664,
080.C-0712,
080.D-0047,
080.D-0086,
080.D-0151,
080.D-0318,
080.D-0347,
080.D-0408,
081.C-0034,
081.C-0119,
081.C-0148,
081.C-0211,
081.C-0388,
081.C-0774,
081.C-0779,
081.C-0802,
081.C-0842,
081.D-0008,
081.D-0065,
081.D-0109,
081.D-0531,
081.D-0610,
081.D-0870,
082.B-0610,
082.C-0040,
082.C-0212,
082.C-0308,
082.C-0312,
082.C-0315,
082.C-0333,
082.C-0357,
082.C-0390,
082.C-0412,
082.C-0427,
082.C-0608,
082.C-0718,
083.C-0186,
083.C-0413,
083.C-0794,
083.C-1001,
083.D-0668,
084.C-0185,
084.C-0228,
084.C-0229,
084.C-1039,
085.C-0019,
085.C-0063,
085.C-0318,
085.C-0393,
086.C-0145,
086.C-0230,
086.C-0284,
086.D-0240,
087.C-0012,
087.C-0368,
087.C-0649,
087.C-0831,
087.C-0990,
087.D-0511,
088.C-0011,
088.C-0323,
088.C-0353,
088.C-0513,
088.C-0662,
089.C-0006,
089.C-0050,
089.C-0151,
089.C-0415,
089.C-0497,
089.C-0732,
089.C-0739,
090.C-0395,
090.C-0421,
090.C-0540,
090.C-0849,
091.C-0034,
091.C-0184,
091.C-0271,
091.C-0438,
091.C-0456,
091.C-0471,
091.C-0844,
091.C-0853,
091.C-0866,
091.C-0936,
091.D-0469,
092.C-0282,
092.C-0454,
092.C-0579,
092.C-0721,
092.C-0832,
092.D-0261,
093.C-0062,
093.C-0409,
093.C-0417,
093.C-0474,
093.C-0919,
094.C-0090,
094.C-0297,
094.C-0428,
094.C-0797,
094.C-0894,
094.C-0901,
094.C-0946,
094.D-0056,
094.D-0596,
095.C-0040,
095.C-0105,
095.C-0367,
095.C-0551,
095.C-0718,
095.C-0799,
095.C-0947,
095.D-0026,
095.D-0717,
096.C-0053,
096.C-0082,
096.C-0183,
096.C-0210,
096.C-0331,
096.C-0417,
096.C-0460,
096.C-0499,
096.C-0657,
096.C-0708,
096.C-0762,
096.C-0876,
096.D-0402,
096.D-0717,
097.C-0021,
097.C-0090,
097.C-0390,
097.C-0434,
097.C-0561,
097.C-0571,
097.C-0864,
097.C-0948,
097.C-1025,
097.D-0156,
097.D-0717,
098.C-0269,
098.C-0292,
098.C-0304,
098.C-0366,
098-C-0518,
098.C-0518,
098.C-0739,
098.C-0820,
098.C-0860,
098.D-0717,
099.C-0093,
099.C-0138,
099.C-0205,
099.C-0303,
099.C-0304,
099.C-0374,
099.C-0458,
099.C-0491,
099.C-0798,
099.C-0880,
099.C-0898,
099.D-0717,
1101.C-0721,
180.C-0886,
183.C-0437,
183.C-0972,
183.D-0729,
184.C-0639,
184.C-0815,
185.D-0056,
188.C-0265,
188.C-0779,
190.C-0027,
191.C-0505,
191.C-0873,
192.C-0224,
192.C-0852,
196.C-0042,
196.C-1006,
198.C-0169,
198.C-0836,
198.C-0838,
281.D-5052,
281.D-5053,
282.C-5034,
282.C-5036,
282.D-5006,
283.C-5017,
283.C-5022,
288.C-5010,
292.C-5004,
295.C-5031,
495.L-0963,
60.A-9036,
60.A-9700, and
63.A-9036.
This research has made use of the SIMBAD database, operated at CDS, Strasbourg, France.
This work has made use of data from the European Space Agency (ESA) mission {\em Gaia} (\url{https://www.cosmos.esa.int/gaia}), processed by
the {\em Gaia} Data Processing and Analysis Consortium (DPAC,
\url{https://www.cosmos.esa.int/web/gaia/dpac/consortium}). Funding
for the DPAC has been provided by national institutions, in particular
the institutions participating in the {\em Gaia} Multilateral Agreement.

\end{acknowledgements}

\bibliographystyle{aa}
\bibliography{nzp_harps}

\appendix

\section{\label{sec:OnlineTables}Online Tables}


\begin{sidewaystable*}
\resizebox{0.72\textheight}{!}
     {\begin{minipage}{1.1\textwidth}
     
    \caption{\label{tab:RVs} RVs and auxiliary data for GJ\,1.  More columns and targets are online available.}
    \centering
    \begin{tabular}{@{}crrrrrrrrrrrrrrrrrrrrrrr@{}}
        \topline
        BJD  &  RV\tablefootmark{a} & $\sigma_{\rm RV}$\tablefootmark{a}
             &  RV\tablefootmark{b} & $\sigma_{\rm RV}$\tablefootmark{b}   
             &  RV\tablefootmark{c} & $\sigma_{\rm RV}$\tablefootmark{c}
             &  RV\tablefootmark{d} & $\sigma_{\rm RV}$\tablefootmark{d} 
             &  RV\tablefootmark{e} & $\sigma_{\rm RV}$\tablefootmark{e}                
             & CRX\tablefootmark{f} & $\sigma_{\rm CRX}$
             & dLW\tablefootmark{g} & $\sigma_{\rm dLW}$
             & H$\alpha$\tablefootmark{h} &  $\sigma_{\rm H\alpha}$ &
             Flag\tablefootmark{i} &
             FWHM\tablefootmark{j} &
             Contrast\tablefootmark{k} &
             BIS-span\tablefootmark{l} &
             RV$_{\rm guess}$\tablefootmark{m} &
             \dots &
             S/N\tablefootmark{n}                          \\
            & [m/s] & [m/s] & [m/s] & [m/s] &  [m/s] & [m/s] & [m/s] & [m/s] & [m/s] & [m/s] & [m/s] & [m/s] & [${\rm m^2/s^2}$]  & [${\rm m^2/s^2}$]& & & & [km/s] &   & [km/s] & [km/s] & &     \\
        \midline
        2452985.59653    &   -2.514    &   1.095  & 25699.819  &   1.210   &	-15.674 &  0.659 &  25700.010 & 0.676 & -15.664 & 0.659 & -7.412  & 6.668 & -6.273  & 1.632 & 0.8970 & 0.0014 & 0  & 3.063 & 22.608 & -0.008  &   22.900 & \dots & 63.2  \\
2452998.57798    &   -4.221    &   0.944  & 25698.307  &   1.060   &	-17.566 &  0.358 &  25698.833 & 0.340 & -17.559 & 0.358 & -6.070  & 4.723 &-11.963  & 1.466 & 0.9050 & 0.0008 & 0  & 3.063 & 22.806 & -0.008  &   22.900 &\dots & 111.0  \\ 
2453206.89682    &   -1.116    &   0.929  & 25702.054  &   1.043   &	-16.049 &  0.316 &  25702.821 & 0.282 & -16.051 & 0.316 & -0.118  & 2.682 & -4.681  & 1.232 & 0.9019 & 0.0006 & 0  & 3.049 & 22.393 & -0.009  &   22.900 &\dots  &139.1  \\ 
2453335.61883    &    1.021    &   0.983  & 25703.493  &   1.096   &	-12.888 &  0.451 &  25707.205 & 0.441 & -12.883 & 0.451 & 10.358  & 3.735 & -5.924  & 0.994 & 0.8859 & 0.0010 & 0  & 3.055 & 22.572 & -0.008  &   22.900 & \dots & 87.2 \\ 
2453520.93453    &    1.126    &   1.308  & 25706.232  &   1.506   &	-11.721 &  0.978 &  25711.854 & 1.172 & -11.714 & 0.978 &  3.504  & 8.350 & -1.273  & 1.467 & 0.8895 & 0.0018 & 0  & 3.048 & 22.369 & -0.013  &   22.900 & \dots & 46.5 \\ 
2453572.93160    &   -0.419    &   0.940  & 25703.017  &   1.048   &	-14.635 &  0.347 &  25707.779 & 0.300 & -14.637 & 0.347 & -3.472  & 3.036 &  5.757  & 0.701 & 0.8835 & 0.0007 & 0  & 3.065 & 22.513 & -0.009  &   22.900 &\dots & 128.6  \\ 
2453575.87158    &   -0.286    &   0.646  & 25702.267  &   0.607   &	-14.599 &  0.268 &  25707.395 & 0.229 & -14.596 & 0.268 & -0.538  & 2.684 & -0.391  & 0.662 & 0.8822 & 0.0005 & 0  & 3.061 & 22.524 & -0.008  &   22.900 &\dots & 168.6  \\ 
2453668.67121    &    2.319    &   0.919  & 25704.765  &   0.971   &	-10.891 &  0.284 &  25711.888 & 0.247 & -10.894 & 0.284 & -0.515  & 2.681 & -0.249  & 0.641 & 0.8879 & 0.0005 & 0  & 3.063 & 22.490 & -0.008  &   22.900 &\dots & 156.0  \\ 
2453672.68499    &    0.067    &   0.912  & 25702.668  &   1.029   &	-13.147 &  0.260 &  25710.106 & 0.224 & -13.149 & 0.260 &  3.784  & 2.308 & -0.964  & 0.695 & 0.8855 & 0.0005 & 0  & 3.065 & 22.512 & -0.008  &   25.200 &\dots & 172.1  \\ 
2453692.60131    &    1.206    &   0.940  & 25702.540  &   0.884   &	-12.623 &  0.574 &  25710.342 & 0.517 & -12.602 & 0.574 & -7.333  & 4.027 & -5.672  & 0.956 & 0.8729 & 0.0011 & 0  & 3.069 & 22.589 & -0.013  &   25.200 & \dots & 76.8  \\ 
2453694.61633    &   -1.026    &   0.955  & 25701.745  &   0.999   &	-13.279 &  0.436 &  25708.826 & 0.379 & -13.269 & 0.436 &  8.826  & 4.143 & -6.577  & 0.836 & 0.8875 & 0.0008 & 0  & 3.062 & 22.525 & -0.006  &   25.200 &\dots & 101.7  \\ 
2453700.60008    &   -1.484    &   0.949  & 25701.340  &   1.058   &	-14.273 &  0.387 &  25708.898 & 0.333 & -14.268 & 0.387 & -1.716  & 3.467 & -6.100  & 0.859 & 0.8808 & 0.0007 & 0  & 3.063 & 22.536 & -0.009  &   25.200 &\dots & 116.0  \\ 
2453721.56438    &    4.497    &   1.087  & 25706.687  &   1.172   &	 -9.075 &  0.647 &  25714.029 & 0.604 &  -9.067 & 0.647 &  6.361  & 5.061 & -5.462  & 0.886 & 0.8900 & 0.0012 & 0  & 3.059 & 22.518 & -0.010  &   25.200 &\dots &  68.7  \\
2454291.92368    &   -3.192    &   0.948  & 25701.836  &   1.025   &	-16.274 &  0.366 &  25713.460 & 0.319 & -16.266 & 0.366 &  1.642  & 3.711 &  6.308  & 0.549 & 0.9041 & 0.0007 & 0  & 3.068 & 22.429 & -0.008  &   25.200 &\dots & 123.5  \\ 
2454295.89151    &   -2.589    &   0.840  & 25700.771  &   0.845   &	-15.789 &  0.324 &  25714.415 & 0.276 & -15.780 & 0.324 & 10.511  & 3.380 & -3.017  & 0.672 & 0.9093 & 0.0006 & 0  & 3.063 & 22.543 & -0.008  &   25.200 &\dots & 141.6  \\ 
2454341.81220    &    0.676    &   0.955  & 25702.759  &   1.058   &	-12.272 &  0.385 &  25716.681 & 0.333 & -12.272 & 0.385 & -9.892  & 3.543 &  8.473  & 0.676 & 0.9077 & 0.0007 & 0  & 3.063 & 22.290 & -0.009  &   25.200 &\dots & 118.5  \\ 
2454343.83867    &    1.829    &   0.955  & 25703.964  &   1.057   &	-11.127 &  0.385 &  25717.855 & 0.332 & -11.128 & 0.385 & -4.562  & 3.837 &  7.147  & 0.880 & 0.9290 & 0.0008 & 0  & 3.074 & 22.428 & -0.011  &   25.200 &\dots & 118.4  \\
2454345.81467    &    0.405    &   0.878  & 25702.794  &   0.849   &	-13.538 &  0.275 &  25715.362 & 0.231 & -13.522 & 0.275 & -3.577  & 2.599 & -2.768  & 0.699 & 0.8962 & 0.0005 & 0  & 3.067 & 22.490 & -0.008  &   25.200 &\dots & 170.8  \\ 
2454346.70406    &    0.249    &   0.887  & 25702.720  &   0.907   &	-11.880 &  0.314 &  25718.167 & 0.269 & -11.885 & 0.314 & -1.458  & 2.623 & -3.511  & 0.677 & 0.8961 & 0.0006 & 0  & 3.070 & 22.553 & -0.009  &   25.200 &\dots & 145.3  \\ 
2454346.80173    &   -0.493    &   0.923  & 25700.768  &   0.935   &	-12.650 &  0.404 &  25716.027 & 0.351 & -12.657 & 0.404 & -1.931  & 4.308 &  4.547  & 0.707 & 0.8984 & 0.0008 & 0  & 3.070 & 22.402 & -0.009  &   25.200 &\dots & 112.3  \\ 
2454390.65910    &    0.798    &   0.889  & 25702.334  &   0.878   &	-10.780 &  0.378 &  25718.359 & 0.328 & -10.773 & 0.378 &  2.921  & 3.675 &  4.241  & 0.503 & 0.8887 & 0.0007 & 0  & 3.067 & 22.388 & -0.009  &   25.200 &\dots & 119.8  \\ 
2454391.63492    &    4.038    &   1.001  & 25706.781  &   1.094   &	 -8.324 &  0.489 &  25721.982 & 0.434 &  -8.316 & 0.489 &  7.090  & 4.264 & -5.475  & 0.653 & 0.8837 & 0.0010 & 0  & 3.073 & 22.670 & -0.009  &   25.200 &\dots &  89.7  \\ 
2454392.68951    &    1.710    &   0.956  & 25704.270  &   1.060   &	-10.667 &  0.387 &  25719.376 & 0.340 & -10.665 & 0.387 &  4.929  & 2.997 &  3.513  & 0.544 & 0.9251 & 0.0008 & 0  & 3.064 & 22.392 & -0.010  &   25.200 &\dots & 115.5  \\ 
2454393.51202    &    1.877    &   0.919  & 25703.512  &   1.034   &	-10.451 &  0.285 &  25718.937 & 0.249 & -10.446 & 0.285 & -4.046  & 2.586 & -2.042  & 0.478 & 0.8776 & 0.0006 & 0  & 3.068 & 22.499 & -0.008  &   25.200 &\dots & 156.3  \\ 
2454393.74815    &    0.860    &   0.912  & 25703.391  &   1.029   &	-11.535 &  0.259 &  25718.360 & 0.227 & -11.535 & 0.259 & -0.065  & 2.245 & -0.680  & 0.484 & 0.8776 & 0.0005 & 0  & 3.067 & 22.478 & -0.009  &   25.200 &\dots & 173.0  \\ 
2454394.71224    &   -0.216    &   0.928  & 25704.183  &   1.043   &	-10.666 &  0.320 &  25719.102 & 0.284 & -10.658 & 0.320 & -4.975  & 2.777 &  0.047  & 0.631 & 0.8766 & 0.0006 & 0  & 3.074 & 22.535 & -0.007  &   25.200 &\dots & 137.8  \\ 
2454421.53195    &   -1.793    &   0.765  & 25701.630  &   1.020   &	-14.626 &  0.243 &  25716.845 & 0.205 & -14.620 & 0.243 &  2.545  & 2.509 & -7.272  & 0.748 & 0.9087 & 0.0005 & 0  & 3.060 & 22.523 & -0.009  &   25.200 &\dots & 190.8  \\
2454425.57119    &   -2.693    &   0.873  & 25702.034  &   0.604   &	-14.025 &  0.337 &  25717.252 & 0.295 & -14.021 & 0.337 &  3.880  & 3.378 & -5.473  & 0.780 & 0.9075 & 0.0007 & 0  & 3.067 & 22.576 & -0.009  &   25.200 &\dots & 132.1  \\ 
2454447.56185    &   -0.762    &   0.700  & 25702.218  &   0.691   &	-13.708 &  0.383 &  25716.783 & 0.335 & -13.710 & 0.383 &  4.801  & 4.114 &  2.729  & 0.568 & 0.8836 & 0.0007 & 0  & 3.060 & 22.331 & -0.008  &   25.200 &\dots & 117.3  \\ 
2454449.55934    &   -1.295    &   0.800  & 25702.376  &   0.856   &	-12.670 &  0.451 &  25718.613 & 0.400 & -12.666 & 0.451 &  0.720  & 3.770 &  1.249  & 0.439 & 0.8820 & 0.0009 & 0  & 3.065 & 22.392 & -0.010  &   25.200 & \dots & 98.0  \\ 
2454451.57231    &    0.448    &   0.796  & 25703.089  &   0.946   &	-11.550 &  0.380 &  25719.254 & 0.334 & -11.546 & 0.380 & -2.825  & 3.162 &  6.892  & 0.529 & 0.8732 & 0.0007 & 0  & 3.065 & 22.331 & -0.008  &   25.200 &\dots & 117.7  \\
2454460.57518    &    3.226    &   0.745  & 25706.105  &   1.069   &	 -9.117 &  0.429 &  25721.908 & 0.380 &  -9.109 & 0.429 & -3.869  & 3.841 & -1.852  & 0.745 & 0.8826 & 0.0008 & 0  & 3.061 & 22.414 & -0.011  &   25.200 &\dots & 103.0  \\ 
2454461.59124    &    4.553    &   2.907  & 25705.445  &   4.502   &	 -8.276 &  2.773 &  25721.137 & 4.389 &  -8.270 & 2.773 & -1.513  &28.264 & 10.657  & 4.217 & 0.8742 & 0.0047 & 0  & 3.063 & 22.252 & -0.009  &   25.200 & \dots & 17.5  \\ 
2454461.62574    &    0.748    &   0.986  & 25702.091  &   1.081   &	-12.091 &  0.456 &  25717.717 & 0.402 & -12.091 & 0.456 &  3.000  & 4.982 &  2.510  & 0.803 & 0.8773 & 0.0009 & 0  & 3.057 & 22.315 & -0.010  &   25.200 &\dots &  97.7  \\ 
2454464.56295    &   -0.032    &   0.770  & 25701.071  &   0.941   &	-13.393 &  0.351 &  25716.452 & 0.310 & -13.386 & 0.351 &  1.772  & 3.335 & -0.059  & 0.512 & 0.8691 & 0.0007 & 0  & 3.065 & 22.440 & -0.009  &   25.200 &\dots & 126.8  \\ 
2454660.91236    &    0.243    &   0.968  & 25702.775  &   0.783   &	-12.591 &  0.437 &  25720.065 & 0.368 & -12.592 & 0.437 &  3.565  & 4.044 &  4.012  & 0.986 & 0.9028 & 0.0008 & 0  & 3.068 & 22.444 & -0.007  &   25.200 &\dots & 107.4  \\ 
2454665.92513    &    0.294    &   0.946  & 25703.210  &   0.968   &	-12.384 &  0.363 &  25720.685 & 0.303 & -12.378 & 0.363 & -4.215  & 3.456 &  7.881  & 0.662 & 0.9062 & 0.0007 & 0  & 3.070 & 22.381 & -0.011  &   25.200 &\dots & 131.3  \\ 
2454672.90246    &    2.718    &   0.482  & 25706.140  &   0.517   &	-10.457 &  0.314 &  25722.745 & 0.262 & -10.444 & 0.314 &  1.393  & 3.132 &  4.171  & 0.451 & 0.8996 & 0.0006 & 0  & 3.074 & 22.482 & -0.008  &   25.200 &\dots & 151.5  \\ 
2454682.86111    &    3.706    &   0.742  & 25706.336  &   0.770   &	 -8.852 &  0.383 &  25724.252 & 0.325 &  -8.851 & 0.383 &  2.117  & 3.865 &  1.422  & 0.926 & 0.9106 & 0.0007 & 0  & 3.074 & 22.529 & -0.008  &   25.200 &\dots & 121.6  \\ 
2454701.85816    &   -4.319    &   0.901  & 25699.381  &   0.735   &	-14.817 &  0.331 &  25718.604 & 0.291 & -14.812 & 0.331 & -2.504  & 2.761 &  1.913  & 0.547 & 0.9061 & 0.0007 & 0  & 3.063 & 22.410 & -0.010  &   25.200 &\dots & 135.6  \\ 
2454705.81497    &   -4.608    &   1.005  & 25698.005  &   1.050   &	-15.166 &  0.608 &  25717.439 & 0.546 & -15.166 & 0.608 & -0.165  & 5.574 &  4.098  & 1.019 & 0.9043 & 0.0012 & 0  & 3.059 & 22.338 & -0.010  &   25.200 & \dots & 75.4  \\ 
2454775.64849    &   -1.932    &   0.961  & 25701.061  &   1.064   &	-13.289 &  0.399 &  25721.381 & 0.353 & -13.280 & 0.399 &  2.940  & 3.500 & -0.008  & 0.820 & 0.8916 & 0.0008 & 0  & 3.066 & 22.487 & -0.006  &   25.200 &\dots & 111.3  \\ 
2454778.56711    &   -1.353    &   0.930  & 25700.486  &   1.041   &	-12.787 &  0.317 &  25721.071 & 0.276 & -12.784 & 0.317 & -2.542  & 3.254 & -0.849  & 0.693 & 0.9493 & 0.0006 & 0  & 3.065 & 22.486 & -0.009  &   25.200 &\dots & 142.4  \\ 
2454825.52534    &   -1.597    &   0.940  & 25700.379  &   0.697   &	-14.208 &  0.346 &  25719.569 & 0.301 & -14.198 & 0.346 &  2.674  & 2.873 &  1.115  & 0.726 & 0.8814 & 0.0007 & 0  & 3.065 & 22.434 & -0.009  &   25.200 &\dots & 131.6  \\ 
2454828.54795    &    0.163    &   1.014  & 25701.971  &   0.956   &	-12.454 &  0.514 &  25722.761 & 0.455 & -12.453 & 0.514 & -4.554  & 4.516 &  1.376  & 0.766 & 0.8794 & 0.0010 & 0  & 3.062 & 22.373 & -0.008  &   25.200 &\dots &  87.4  \\ 
2455041.78317    &    0.209    &   0.931  & 25702.846  &   1.042   &	-13.064 &  0.320 &  25723.760 & 0.280 & -13.059 & 0.320 &  0.523  & 2.943 &  4.418  & 0.643 & 0.8878 & 0.0006 & 0  & 3.066 & 22.409 & -0.008  &   25.200 &\dots & 142.0  \\ 
2455044.87569    &    0.040    &   0.941  & 25702.625  &   1.050   &	-13.233 &  0.348 &  25723.559 & 0.307 & -13.230 & 0.348 & -0.040  & 2.768 &  6.945  & 0.789 & 0.8892 & 0.0007 & 0  & 3.069 & 22.409 & -0.010  &   25.200 &\dots & 129.7  \\ 
2455048.82893    &   -2.113    &   0.645  & 25701.114  &   0.696   &	-14.818 &  0.381 &  25722.147 & 0.337 & -14.815 & 0.381 & -3.379  & 3.238 &  4.620  & 0.810 & 0.8847 & 0.0008 & 0  & 3.070 & 22.411 & -0.009  &   25.200 &\dots & 118.1  \\

        \bottomline
    \end{tabular}
    \tablefoot{
        \tablefoottext{a}{Differential RV from SERVAL corrected for BERV, SA, drift, NZP, NZPdrift.}
        \tablefoottext{b}{RV from DRS corrected for BERV, SA, drift, NZP, NZPdrift.}  
        \tablefoottext{c}{Differential RV from SERVAL corrected for BERV, SA, drift, (non-NZP corrected).}
        \tablefoottext{d}{RV from DRS corrected for BERV, SA, drift, (non-NZP corrected).}   
        \tablefoottext{e}{Differential RV from SERVAL corrected for BERV, SA, drift, (non-NZP corrected, joint RV calculation with -pre and -post fibre upgrade data.)}
        \tablefoottext{f}{Chromatic index from SERVAL.}
        \tablefoottext{g}{Differential line width from SERVAL.}
        \tablefoottext{h}{H$\alpha$ index from SERVAL.}
        \tablefoottext{i}{The SERVAL flag is bitwise: 
        0 - normal (reliable) spectra,
        1 - nosci frame, e.g. calibration files,
    2 - spectra taken in "I$_{\rm 2}$" mode,
    4 - spectra taken in "eggs" mode,
    16 - coordinates too much off,
    32 - spectra not within a nautical twilight (daytime),
    64 - spectra with too low S/N,
    128 - spectra with too high S/N.
        }
        \tablefoottext{j}{CCF full width half maximum measurement from DRS.}
        \tablefoottext{k}{Contrast measurement from DRS.}
        \tablefoottext{l}{Bisector-span measurement from DRS.}
        \tablefoottext{m}{User guess for absolute RV from DRS.}
        \tablefoottext{n}{Signal-to-noise ratio in order 55.}        
    } 
\end{minipage}}

\end{sidewaystable*}

\end{document}